\documentclass[aps,preprint,nofootinbib,preprintnumbers,eqsecnum]{revtex4-1}

\usepackage[usenames, dvipsnames]{color}

\newcommand{\nb}[1]{\color{blue}}

\usepackage[
	  pagebackref=false,
	  colorlinks=true,
      linkcolor=blue,
      urlcolor=blue,
      filecolor=black,
      citecolor=red,
      pdfstartview=FitV,
      pdftitle={},
        pdfauthor={},
        pdfsubject={},
        pdfkeywords={},
        pdfpagemode=None,
        bookmarksopen=true
      ]{hyperref}

\usepackage[normalem]{ulem}
\usepackage{amsmath}
\usepackage{enumerate}
\usepackage{amsfonts}
\usepackage{epsfig}
\usepackage{mathbbol}

\usepackage{amssymb}

\newcommand\half{{\ensuremath{\frac{1}{2}}}}

\newcommand\ket[1]{\ensuremath{\lvert{#1}\rangle}}
\newcommand\bra[1]{\ensuremath{\langle{#1}\rvert}}

\newcommand{\be}{\begin{equation}}
\newcommand{\ee}{\end{equation}}
\newcommand{\bea}{\begin{eqnarray}}
\newcommand{\eea}{\end{eqnarray}}
\newcommand{\bega}{\begin{gather}}
\newcommand{\eega}{\end{gather}}

\newcommand{\bi}{\begin{itemize}}
\newcommand{\ei}{\end{itemize}}
\newcommand{\ben}{\begin{enumerate}}
\newcommand{\een}{\end{enumerate}}
\newcommand{\bca}{\begin{cases}}
\newcommand{\eca}{\end{cases}}
\newcommand{\bln}{\begin{align}}
\newcommand{\eln}{\end{align}}
\newcommand{\bst}{\begin{split}}
\newcommand{\est}{\end{split}}
\def\ie{\begin{equation}\begin{aligned}}
\def\fe{\end{aligned}\end{equation}}
\newcommand{\bma}{\le(\begin{matrix}}
\newcommand{\ema}{\end{matrix}\ri)}

\def\b{{\beta}}
\newcommand\ep{\epsilon}

\newcommand\Sig{\Sigma}

\newcommand\Lam{\Lambda}

\newcommand\Om{\Omega}

\newcommand\ga{{\ensuremath{{\gamma}}}}

\newcommand\De{{\ensuremath{{\Delta}}}}

\newcommand\ov{\over}
\newcommand\ha{{\half}}

\def\le{\left}
\def\ri{\right}

\newcommand\sA{{\ensuremath{{\mathcal A}}}}
\newcommand\sB{{\ensuremath{{\mathcal B}}}}

\newcommand\sH{{\ensuremath{{\mathcal H}}}}

\newcommand\sM{{\ensuremath{{\mathcal M}}}}
\newcommand\sN{{\ensuremath{{\mathcal N}}}}
\newcommand\sO{{\ensuremath{{\mathcal O}}}}

\newcommand\sS{{\mathcal S}}

\newcommand\sX{{\mathcal X}}
\newcommand\sY{{\mathcal Y}}

\newcommand\vx{{\vec x}}
\newcommand\vy{{\vec y}}

\newcommand{\fa}{{\mathfrak a}}
\newcommand{\fb}{{\mathfrak{b}}}
\newcommand{\fc}{{\mathfrak{c}}}

\newcommand{\fr}{{\mathfrak r}}

\newcommand{\bid}{\mathbf{1}}
\newcommand{\pt}{\partial}

\newcommand{\wt}{\widetilde}

\usepackage{graphicx}
\usepackage{caption}
\usepackage{subcaption}
\usepackage{mwe}

\begin{document}

\title{Superadditivity in large $N$ field theories and performance of quantum tasks
}

\preprint{MIT-CTP/5382}

\author{Sam Leutheusser}
\affiliation{ School of Natural Sciences, Institute for Advanced Study,
Princeton, NJ 08540, USA \\
Princeton Gravity Initiative, 
Princeton University, 
Princeton, NJ 08544, USA }

\author{Hong Liu}
\affiliation{Center for Theoretical Physics, 
Massachusetts
Institute of Technology, \\
77 Massachusetts Ave.,  Cambridge, MA 02139 }

\begin{abstract}

 \noindent 

Field theories exhibit dramatic changes in the structure of their operator algebras in the limit where the number of local degrees of freedom ($N$) becomes infinite. An important example of this is that the algebras associated to local subregions may not be additively generated in the limit. We investigate examples and explore the consequences of this ``superadditivity'' phenomenon in 
large $N$ field theories and holographic systems.
In holographic examples we find cases in which superadditive algebras can probe the black hole interior, while the additive algebra cannot.
We also discuss how superaddivity explains the sucess of quantum error correction models of holography.
Finally we demonstrate how superadditivity is intimately related to the ability of holographic field theories to perform quantum tasks that would naievely be impossible. We argue that the connected wedge theorems (CWTs) of May, Penington, Sorce, and Yoshida, which characterize holographic protocols for quantum tasks, can be re-phrased in terms of superadditive algebras and use this re-phrasing to conjecture a generalization of the CWTs that is an equivalence statement.

\end{abstract}

\today

\maketitle

\tableofcontents

\section{Introduction}
It has been a long-standing goal to understand the emergence of {\it local} bulk physics in asymptotically AdS spacetimes from holographic conformal field theories~\cite{Maldacena:1997re, Gubser:1998bc, Witten:1998qj}. 

Soon after the discovery of the AdS/CFT duality, it was argued that local physics in wedge subregions of the bulk could only be consistent with the duality if the boundary theory was given by a theory of generalized fields which violates the time-slice axiom\footnote{The time-slice axiom states that all operators on the system can be expressed in terms of those (suitably smeared) on a single Cauchy slice.}~\cite{Duetsch:2002hc}. Indeed a generalized free field theory naturally arises in the large $N$ limit of a holographic CFT, which corresponds to the regime 
where the bulk gravity is semiclassical ($G_N \to 0$).

Recent insights into the behavior of the boundary Hilbert space and operator algebras in the large $N$ limit have shed new light on the emergence of local bulk physics and bulk spacetime~\cite{shortPaper, longPaper, subalgSubreg, Netta,Gesteau:2024rpt}. 
These findings elucidated how the bulk geometric and physical properties such as local physics in a subregion, spacetime connectivity, existence of a horizon, causal structure, and different notions of time, emerge in the boundary theory.

The violation of the time-slice axiom is not the only exotic property required of the boundary theory by AdS/CFT duality. It was pointed out in~\cite{Casini:2019kex} that entanglement wedge reconstruction implies that certain additivity properties of local QFT must be violated.\footnote{The violation of additivity in the boundary theory for AdS/CFT about the vacuum state has been known since very soon after its discovery~\cite{Rehren:1999jn}; however, such violations in more general settings have only been understood recently.} 
The additivity property of a QFT relates the operator algebra of a large subregion to those of smaller subregions whose union covers the large subregion. More explicitly, consider topologically trivial open subregions $R_1, R_2,R_1 \cup R_2$ of a Cauchy slice and the algebras $\sB_{R_1}, \sB_{R_2}, \sB_{R_1 \cup R_2},$ of operators respectively localized in them. The additivity property is the statement
\be \label{eq:additivity}
	\sB_{R_1 \cup R_2} = \sB_{R_1} \vee \sB_{R_2} \equiv \le(\sB_{R_1} \cup \sB_{R_2}\ri)'' \ ,
\ee
where $'$ denotes the commutant operation. In other words, the algebras associated to topologically trivial spacetime subregions are locally generated; we do {\it not} get more operators by considering the algebra $\sB_{R_1 \cup R_2}$ than 
the one generated by the union of $\sB_{R_1}$ and $\sB_{R_2}$. 
Equation~\eqref{eq:additivity} can be violated for topologically nontrivial regions by Wilson lines, or in the presence of generalized symmetries~\cite{Casini:2021zgr}. Theories with such additivity violations arising from generalized symmetries have been called ``incomplete''.

In~\cite{subalgSubreg}, we argued that a holographic CFT that obeys the additivity property at finite $N$, can
develop a superadditivity property in the large $N$ limit. More explicitly, denote the large $N$ limit of the algebras of 
regions $R_1, R_2,$ and $R_1 \cup R_2$ respectively as $\sX_{R_1}, \sX_{R_2},$ and $\sX_{R_1 \cup R_2}$. We then have 
\be \label{supadd}
\sX_{R_1} \vee \sX_{R_2} \subsetneq \sX_{R_1 \cup R_2} \ .
\ee 
Equation~\eqref{supadd} says that we {\it can} get more operators by considering the algebra $\sX_{R_1 \cup R_2}$ than 
the one generated by the union of $\sX_{R_1}$ and $\sX_{R_2}$. In other words, the algebras are {\it not locally} generated even for topologically trivial regions. The property~\eqref{supadd} 
has been used in~\cite{subalgSubreg} to explain various features of entanglement wedge reconstruction, in particular, the quantum error correcting properties~\cite{Almheiri:2014lwa}. 

 In this paper, we first illustrate further the superadditivity property~\eqref{supadd} of a holographic CFT with more examples, both from the boundary perspective and from requirements of the duality with the bulk. In particular, we highlight that superadditivity leads to entanglement wedge nesting in the bulk. 
 We then discuss some implications of this property for the bulk system. These include an explanation of the quantum error correcting properties of the bulk and new results related to holographic quantum tasks.
Specifically, we show that the superadditivity property enables holographic CFTs to perform quantum tasks non-locally, and we propose a reformation and generalization of the connected wedge theorems of~\cite{May:2019yxi, May:2019odp, May:2021nrl, May:2022clu}. 

We also demonstrate how superadditive algebras related to certain boundary subregions can see behind the horizons of black holes even when additive algebras cannot. Specifically in the context of the duality between the BTZ black hole and two copies of a holographic CFT$_2$ on $\mathbb{R}\times S^1$ we consider having access to the entire `left' CFT and an angular interval of half-width $w$ in the `right' CFT. For small values of $w$ neither the additive nor superadditive algebra can describe the black hole interior, while for large values of $w$ both algebras can describe this physics. However in an intermediate regime $\pi - w_c < w < w_c$, we find that the superadditive algebra probes the interior while the additive algebra only probes the exterior. In the high-temperature limit one finds that the critical value behaves as $w_c \to \pi$ demonstrating the superadditivity is extremely pronounced: the additive algebra never sees the interior while the superadditive algebra always does.

The plan of the paper is as follows. In section~\ref{sec:addAnom} we discuss the phenomenon of superadditivity from a general perspective and highlight how it can arise in large $N$ systems. We then discuss its manifestation in holographic systems, followed by several examples both in field theory and holography. We then turn to the implications of superadditivity. In section~\ref{sec:impQEC} we discuss how superadditivity explains the observed quantum error correcting~(QEC) properties of holographic duality. We then discuss to what extent we can understand entanglement wedge reconstruction at a finite value of $N$ (finite $G_N$) and in what sense superadditivity persists in such a non-perturbative treatment. We also comment on the features both captured and missed by QEC models of holography. In section~\ref{sec:cwt} we discuss the implications of superaddivity for holographic quantum tasks. We first review the discussions of the connected wedge theorems (CWTs) of~\cite{May:2019yxi, May:2019odp, May:2021nrl, May:2022clu} and then re-interpret these theorems in terms of superadditivity. Using this new interpretation, we conjecture a generalization of the CWTs and provide some preliminary arguments in support of this conjecture. Finally, we conclude in section~\ref{sec:disc} with a discussion of what we have learned from the superadditivity property of large $N$ field theories and some open questions concerning this property.

\medskip
\noindent {\bf Notations and conventions} 

\medskip 
Here we list our notations, definitions, and assumptions about the bulk and boundary theories.

\bi

\item We use captial latin letters, e.g. $R,~S$ for boundary spatial (or spacetime) subregions, lowercase gothic letters, e.g. $\fa,~\fb$ for bulk spatial (or spacetime) subregions. 

\item A Cauchy slice of the entanglement wedge of a boundary subregion $R$ (i.e. a homology hypersurface when $R$ is a spatial boundary subregion) is denoted by $\fb_R.$ 

\item The causal domain of $R$ is denoted by $\hat \fc_R$, which is defined as the causal completion of the causal wedge of $R$. 

\item  Causal complements are denoted with a prime and are taken in the spacetime in which the set is originally defined, i.e. $R'$ is the boundary causal complement of $R$ and $\fa'$ is the bulk causal complement of $\fa.$

\item We use hats to denote the domain of dependence in the spacetime the set is originally defined, i.e. $\hat \fa$ is the domain of dependence of $\fa$ in the bulk geometry. 

\item For spatial subregions, the complement on a single Cauchy slice is denoted with a bar, i.e. $\bar{R}$ is the complement of a spatial boundary subregion $R$ on a boundary Cauchy slice. 

\item The (weak closure of the) algebra generated by bulk fields in a bulk subregion $\fa$ is denoted by $\widetilde{\sM}_{\fa}.$ 

\item We denote the algebra generated by single-trace operators in an open region $A$ by $\sY_A$.

\item For $R$ a spatial boundary subregion, the boundary algebra dual to the causal domain is denoted by $\sY_{\hat R}$, 
while the algebra dual to the entanglement wedge is denoted by $\sX_R.$

\ei

\section{Superadditivity for large $N$ systems} \label{sec:addAnom}

\subsection{General description}

Consider a relativistic quantum field in a globally hyperbolic spacetime. For each open subregion $R$ we can associate a von Neumann algebra $\sB_R$ of  physical observables in $R$. 
The collection $\{\sB_R\}$ for all possible choices of $R$ satisfies the relations~\cite{HaagKastler, Haag:1992hx}
\bea 
\label{he1}
	&\sB_{R_1} \subseteq \sB_{R_2}~\text{if}~R_1 \subseteq R_2 , \\
	\label{Hev}
	&\sB_{\hat R} = \sB_{R} , \\
	\label{Hev1}
	&\sB_{R'} \subseteq \le(\sB_{R}\ri)' \ .
\eea

We now restrict to the case where $R_1$ and $R_2$ are spatial subregions of a common Cauchy slice. Equation~\eqref{he1} is a statement of locality, i.e. if a region $R_1$ is contained in $R_2$ then all operations that can be performed in $R_1$ should be a subset of those in $R_2$. For example, it implies
\be
\label{Hev2}
	\sB_{R_1} \lor \sB_{R_2} \equiv \le(\sB_{R_1} \cup \sB_{R_2}\ri)'' \subseteq \sB_{R_1 \cup R_2}  \ .
\ee
From~\eqref{he1}, both $\sB_{R_1}$ and $\sB_{R_2}$ are contained in $\sB_{R_1 \cup R_2}$. The left hand side (LHS) 
of~\eqref{Hev2} is the smallest von Neumann algebra containing both $\sB_{R_1}$ and $\sB_{R_2}$, and thus~\eqref{Hev2} follows. 

 Equation~\eqref{Hev} is a consequence of the causal nature of the equations of motion, i.e. we can express an operator in $\hat R$ in terms of those in $R$ via equations of motion. In particular, it means that the entire operator content of the theory may be described within a neighborhood of a single Cauchy slice. Equation~\eqref{Hev1} is another statement of causality: operators in the causal complement of $R$, which we denote by $R',$ should commute with those in $R$.

The properties~\eqref{he1}--\eqref{Hev1} are believed to be satisfied by any relativistic quantum field theory that has local equations of motion. However, there are additional properties that are expected to be satisfied by so-called `complete' theories (see e.g. a discussion in~\cite{Witten:2023qsv}). 
The first of these is known as Haag duality, which says that the causality condition~\eqref{Hev1} is saturated
\be \label{eq:haagDuality}
	\sB_{R'} = \le(\sB_R\ri)' . \
\ee 
Another possible property is the saturation of equation~\eqref{Hev2} 
\be \label{additivity}
	\sB_{R_1 \cup R_2} = \sB_{R_1} \vee \sB_{R_2} \ ,
\ee
in which case we say that the algebras are generated additively. 

Both equations~\eqref{eq:haagDuality} and~\eqref{additivity} imply a stronger form of locality. Equation~\eqref{eq:haagDuality} says that the only operators that commute with those in $\sB_R$ lie in $R'$, while~\eqref{additivity} says that there are no additional operators in a joint subregion that cannot separately be expressed in terms of~(limits of) sums of products of operators in the individual subregions making up the union. If both~\eqref{eq:haagDuality} and~\eqref{additivity} are satisfied, by taking the commutant of~\eqref{additivity}, we also have
\be \label{add1}
	\sB_{R_1 \cap R_2} = \sB_{R_1} \land \sB_{R_2} \equiv \sB_{R_1} \cap \sB_{R_2} \ .
\ee

Generically, the Haag duality and additivity properties~\eqref{eq:haagDuality} and~\eqref{additivity} are believed to hold for topologically trivial regions. We will be interested in such theories that are also defined with a parameter characterizing the number of local field theoretic degrees of freedom, denoted by $N,$ that can be dialed. 

Even though each finite $N$ theory may satisfy Haag duality and additivity for topologically trivial regions, these may be violated in the $N \to \infty$ limit. This occurs because, in the large $N$ limit, both the operator algebra and Hilbert space can undergo dramatic changes, as many operators and states may not have a well-defined large $N$ limit and thus drop out.

We say an operator has a sensible large $N$ limit if its vacuum correlation functions have a well-defined $N \to \infty$ limit.  For the $\sN =4$ SYM theory these are finite products of single-trace operators.
They generate a $C^*$-algebra with norm descended from the finite $N$ theories.
We say a state $\ket{\Psi}$ has a sensible large $N$ limit if correlation functions of single-trace operators~(with their expectation values subtracted) in the state have a well-defined $N \to \infty$ limit.

For a general state $\ket{\Psi}$ that survives the large $N$ limit, we define the set of operators that survive the large $N$ limit in $\ket{\Psi}$ as $\sA_\Psi$. By definition of $\ket{\Psi}$, $\sA_\Psi$ includes operators generated by single-trace operators, but may also include operators that have a sensible large $N$ limit only in $\ket{\Psi}$.\footnote{That is, they have well defined correlation functions in $\ket{\Psi}$, but not in other states.}
$\sA_\Psi$ again forms a $C^*$-algebra. 
We can build a Hilbert space  $\sH_{\Psi}^{\rm (GNS)}$ around  $\ket{\Psi}$ by acting elements of the algebra $\sA_\Psi$ on it, through the GNS procedure~\cite{gelfandNaimark, segal}. This Hilbert space gives rise to a representation $\pi_\Psi: \sA_{\Psi} \to \sB\le(\sH_{\Psi}^{\rm (GNS)}\ri).$

Consider a subregion $R$ on a Cauchy slice and denote the complement on the Cauchy slice as $\bar{R}$. At a finite $N$, there is a state-independent algebra of operators in $R$, denoted by $\sB^{(N)}_R$. 
We denote its large $N$ limit as~\cite{subalgSubreg}
\bega \label{xr}
\sX_R = \pi_\Psi \le( \lim_{N \to \infty, \ket{\Psi}} \sB_R^{(N)} \ri)'' ,
\end{gather} 
where we have stressed that the limit is $\ket{\Psi}$-dependent. From~\eqref{Hev} we have 
\be
\sX_R = \sX_{\hat R} \ .
\ee
Another natural object associated with $R$ is 
\be 
\sY_{\hat R} 
\equiv (\pi_\Psi( \sS_{\hat R}))'' \ ,
\ee
where $\sS_{\hat R}$ denotes the set of single-trace operators localized in $\hat R$. 
Below we will always use $\sY_O$ to denote the algebra {generated by} single-trace operators in a region $O$ acting on the GNS Hilbert space. 

Since single-trace operators always survive the large $N$ limit, we have 
\be \label{ehb}
\sY_{\hat R} \subseteq \sX_R \ .
\ee
For a general region $R$, $\sY_{\hat R}$ is a proper subset of $\sX_{R}$; however, in some special cases, for example, if $R$ is a spherical region and $\ket{\Psi}$ is the vacuum state, we have 
$\sY_{\hat R} = \sX_R$.

\begin{figure}[h]
        \centering
        \begin{subfigure}[b]{0.45\textwidth}
            \centering
		\includegraphics[width=\textwidth]{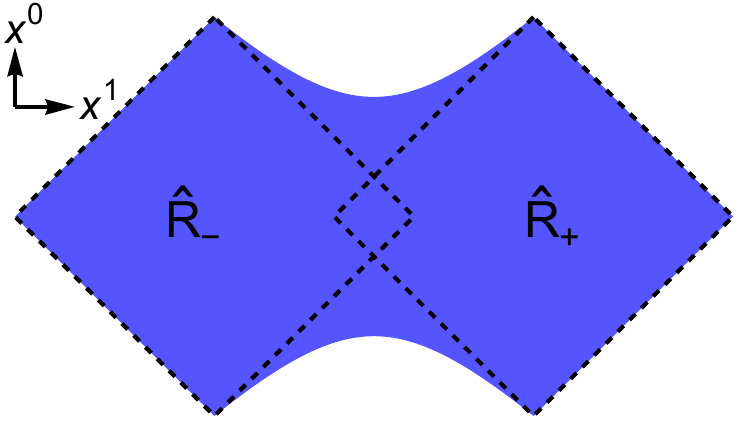} 
            \caption[]%
            {{\small} {Single-trace operators in the region $\hat R_- \cup \hat R_+$ (the regions bounded by the dashed lines in the figure) generate the same algebra as single-trace operators in the entire blue shaded region, $A$.}  }    
        \end{subfigure}
        \hfill
        \begin{subfigure}[b]{0.45\textwidth}   
            \centering 
		\includegraphics[width=0.8\textwidth]{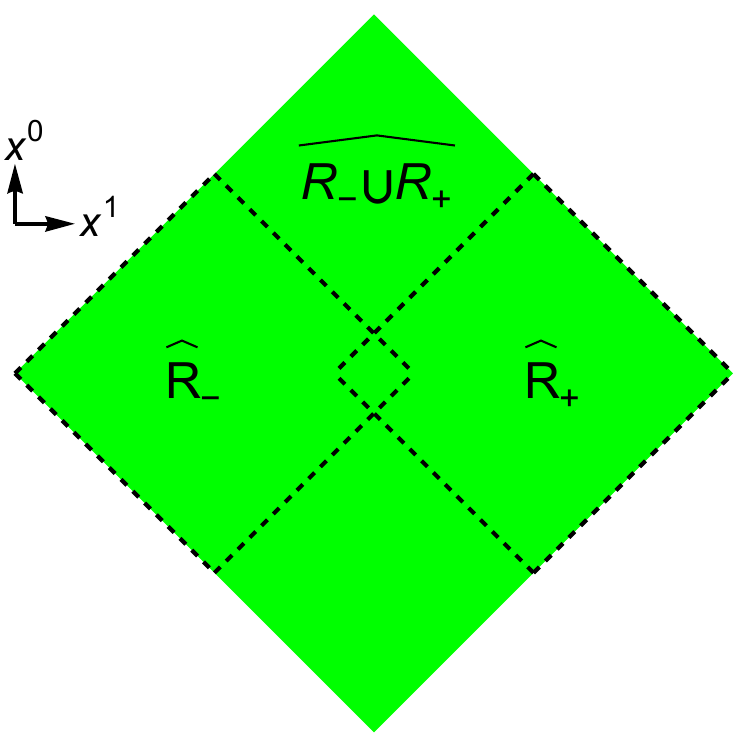}
            \caption[]%
            {{\small   Causal domains of dependence for $R_+, R_-$ and $R_+ \cup R_-$, respectively.
            }}    
        \end{subfigure}
        \caption[  ]
        {\small  About the vacuum state of a CFT$_2$ the algebra $\sX_{R_-}\vee\sX_{R_+}$ is equal to the algebra generated by single-trace operators in the blue region depicted in (a) (see Sec.~\ref{sec:2din} for more detail). In contrast, $\sX_{R_- \cup R_+}$ is equal to the algebra generated by single-trace operators in the entire green region of (b). Clearly $\sX_{R_-} \lor \sX_{R_+} \subsetneq \sX_{R_- \cup R_+}$.
      } 
\label{fig:overint}
\end{figure}

Given that the definition~\eqref{xr} of $\sX_R$ involves a limit, it is natural to ask if Haag duality~\eqref{eq:haagDuality} or additivity~\eqref{additivity} is preserved in the limit. It turns out that the additivity property is violated\footnote{{For previous discussion of such violations in the holographic context see~\cite{Rehren:1999jn, Casini:2019kex, Faulkner:2020hzi}.}} in the limit for certain choices of $R$'s~\cite{subalgSubreg}, i.e. 
\bega \label{hen3}
\sX_{R_1} \lor \sX_{R_2} \subsetneq \sX_{R_1 \cup R_2} .
\end{gather} 
The superadditivity property~\eqref{hen3} means that in taking the large $N$ limit to get $\sX_{R}$, some level of ``nonlocality'' has developed.  See Fig.~\ref{fig:overint} for an explicit example of~\eqref{hen3}.

There is currently no explicit example of a violation of Haag duality {for algebras defined as in~\eqref{xr}}. Below we will see that in holographic systems under reasonable assumptions, Haag duality  is satisfied, 
 \be \label{hdt}
	\sX_{\bar{R}} = \sX'_{R} \ .
\ee 
Taking the commutant on both sides of~\eqref{hen3} and using~\eqref{hdt}, we  have for certain $R$'s
\be
\label{intViol}
	\sX_{\bar R_1 \cap \bar R_2}   \subsetneq  \sX_{\bar R_1} \land \sX_{\bar R_2} \ .
\ee

A special case in which the algebra is generated additively even in the large $N$ limit occurs for $R_2 = \bar R_1$. In this case, assuming the local algebras are factors, we have 
\be 
\sX_{R} \vee \sX_{\bar R} = \sX_{R} \vee  \sX_R' = \sB (\sH_{\Psi}^{\rm (GNS)})= \sX_{R \cup \bar R} \ .
\ee
This example also indicates that a large amount of mutual information is not sufficient for superadditivity.

Finally, we note that the $\sY_{\hat R}$ algebras satisfy only a set-theoretic version of additivity, $\sY_{\hat R_1} \vee \sY_{\hat R_2} = \sY_{\hat R_1\cup \hat R_2},$ and do not satisfy Haag duality.

\subsection{Superadditivity in holographic systems} \label{sec:qec}

Now consider a CFT with a holographic dual, where the parameter $N$ is mapped to $1/G_N$ on the gravity side, and 
the large $N$ limit corresponds to the semi-classical $G_N \to 0$ limit. 
We consider the boundary theory in a state---which we refer to as a semi-classical state---that in the large $N$ limit is described by a classical geometry on the gravity side. 
The bulk theory is described by a quantum field theory in a curved spacetime at leading order in the small $G_N$ and small $\alpha'$ expansion. 
In $1/N$ perturbation theory, the full CFT Hilbert space separates into disjoint GNS Hilbert spaces of small excitations around semi-classical states, $\sH^{\rm (GNS)}_{\Psi}$, each of which is equivalent to the Fock space, $\sH^{\rm (Fock)}_{\Psi},$ of bulk fields quantized on the corresponding classical background geometry in $G_N$ perturbation theory.
{Since the bulk theory arises} from the low energy limit of a consistent quantum gravity theory, {we assume that it is} a complete theory,\footnote{{See~\cite{Witten:2023qsv} for a discussion on completeness of theories in the low energy limit of quantum gravity.}} and thus that additivity and Haag duality are satisfied by its algebras.

It was proposed in~\cite{subalgSubreg} that there is a correspondence between bulk subregions and boundary subalgebras, called subregion-subalgebra duality.  
The duality includes entanglement wedge reconstruction as a sub-case and provides an algebraic description of it. More explicitly, for a subregion $R$ of a boundary Cauchy slice, $\sX_R$ and $\sY_{\hat R}$ discussed above can be identified respectively with the bulk operator algebras associated to the entanglement wedge  and causal domain  of $R$.
Recall that the entanglement wedge is defined as the bulk domain of dependence $\hat \fb_R$ of a region $\fb_R$ 
bounded by the RT surface for $R$ and $R$ itself. The causal domain is defined as the bulk causal completion of the intersection of the bulk past and future of $\hat{R}$. We denote it using its intersection with a certain bulk Cauchy slice, $\fc_R,$ so that the causal domain is $\hat \fc_R.$

With the algebra associated to a bulk region $\mathfrak{o}$ denoted as $\wt \sM_{\mathfrak{o}}$, we then have 
\bega \label{ewr}
\sX_R = \wt \sM_{\fb_R} , \\
\sY_{\hat R} = \wt \sM_{\fc_R} \ .
\label{cwr} 
\end{gather}
Equation~\eqref{ewr} is the statement of entanglement wedge reconstruction and~\eqref{cwr} is sometimes called HKLL reconstruction~\cite{Hamilton:2006az}.

\begin{figure}[!h]
        \centering
        \begin{subfigure}[b]{0.45\textwidth}
            \centering
		\includegraphics[width=\textwidth]{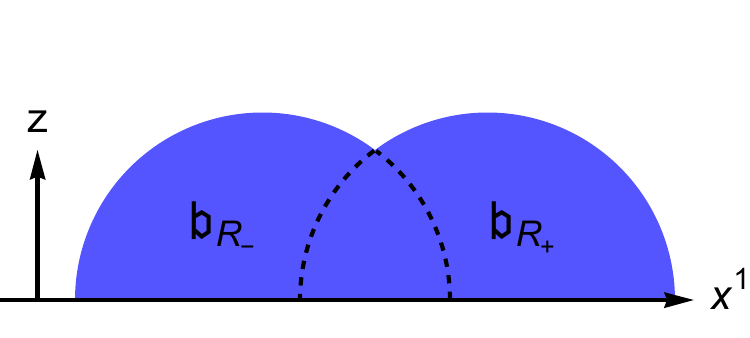} 
            \caption[]%
            {{\small} The bulk dual of $\sX_{R_-}\vee\sX_{R_+}$ is the (causal completion of the) entire blue region, $\fb_{R_-}\cup\fb_{R_+}$. }    
        \end{subfigure}
        \hfill
        \begin{subfigure}[b]{0.45\textwidth}   
            \centering 
		\includegraphics[width=\textwidth]{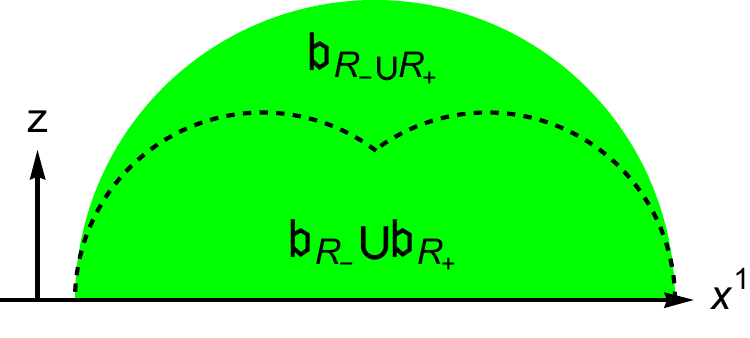}
            \caption[]%
            {{\small The bulk dual of $\sX_{R_- \cup R_+}$ is the (causal completion of the) entire green half-disk, $\fb_{R_- \cup R_+}.$}}    
        \end{subfigure}
        \caption[  ]
        {\small $x^0=0$ slice of Poincare AdS$_3$ is shown in the plots. The proper inclusion $\fb_{R_-}\cup\fb_{R_+} \subsetneq \fb_{R_- \cup R_+}$ is the bulk geometric realization of the superadditivity $\sX_{R_-}\vee\sX_{R_+} \subsetneq \sX_{R_- \cup R_+}.$   } 
\label{fig:bulkint}
\end{figure}

Properties~\eqref{hen3}--\eqref{intViol} have simple geometric interpretations in the bulk. 
From~\eqref{ewr} we have 
\be 
\sX_{R_1} \lor \sX_{R_2} = \wt \sM_{\fb_{R_1}} \lor  \wt \sM_{\fb_{R_2}} = 
\wt \sM_{\fb_{R_1} \cup \fb_{R_2}}, \quad \sX_{R_1 \cup R_2} = \wt \sM_{ \fb_{R_1 \cup R_2}},
\ee
where in the second equality of the first equation we have used that the bulk algebras are additive. 
Equation~\eqref{hen3} then gives 
\be \label{Aoo2}
\fb_{R_1} \cup \fb_{R_2} \subseteq \fb_{R_1 \cup R_2} , 
\ee
i.e. the union of entanglement wedges for two boundary regions $R_1, R_2$ is smaller than that for $R_1 \cup R_2$. 
In Fig.~\ref{fig:bulkint}, we give the bulk dual of the example of Fig.~\ref{fig:overint}.
In the bulk, equation~\eqref{Aoo2} is the statement of entanglement wedge nesting, namely, that $R_1 \subseteq R_2 \Rightarrow \fb_{R_1} \subseteq \fb_{R_2}$. 
Conversely, for holographic systems, (proper) entanglement wedge nesting~\eqref{Aoo2} implies that 
the algebras dual to the bulk entanglement wedges 
are superadditive~\eqref{hen3}. 
Similarly,~\eqref{intViol} is equivalent to the statement that the intersection of the entanglement wedges for $R_1, R_2$ is larger than that of $R_1 \cap R_2$, i.e. 
\be 
\fb_{R_1 \cap R_2} \subseteq \fb_{R_1} \cap \fb_{R_2} \ .
\ee
Haag duality of the bulk theory means that 
\be \label{eq:bulkHaagDuality}
\le(\wt{\sM}_{\fb_R}\ri)' = \wt \sM_{\overline{\fb_R}} 
\ee
where $\overline{\fb_R}$ denotes the complement of $\fb_R$ on a bulk Cauchy slice containing $\fb_R$. 
If the system is in a pure state,  $R$ and $\bar R$ share the same RT/HRT surface, i.e. 
\be \label{eq:sharedRTinPure}
\overline{\fb_R} = \fb_{\bar R} ,
\ee 
which along with~\eqref{eq:bulkHaagDuality} and~\eqref{ewr} implies that 
\be 
\sX_R' =\sX_{\bar R}  \ .
\ee
We therefore see that the boundary Haag duality property given in~\eqref{hdt} is equivalent to bulk Haag duality for bulk subregions that are the entanglement wedge of some boundary subregion.

\subsection{Explicit examples of the superadditivity property} \label{sec:examples}

We now discuss some explicit examples of the superadditivity property~\eqref{hen3}. We first 
give more details for the field theory example of Fig.~\ref{fig:overint}, and discuss its higher dimensional generalizations. For more complicated examples, it becomes difficult to derive the superadditivity property from the field theory side. Fortunately, through holographic duality many examples can be found with the help of the bulk. We discuss several classes of examples below.

\subsubsection{Two intervals in the vacuum of CFT$_2$} \label{sec:2din}

Consider the example of Fig.~\ref{fig:overint}, where two intervals $R_\pm,$ each of half-width $a$, are respectively centered at $x^1 = \pm b$. For $b < a$, the two intervals overlap. 
We would like to compare  $ \sX_{R_-} \lor \sX_{R_+}$ with $ \sX_{R_- \cup R_+}$ in the vacuum state of a CFT$_2$  in the large central charge limit. Note that, for a single interval $R$, $\sX_R$ is simply generated by {(the representations of)} all the single-trace operators in $\hat R$, i.e. $\sX_{R} =  \sY_{\hat R}$.

Since single-trace operators at different times are independent, we might have naively expected $ \sX_{R_-} \lor \sX_{R_+}$ to be generated by single-trace operators localized in $\hat R_- \cup \hat R_+$. The result 
 turns out to be somewhat intricate, but fortunately can be understood using techniques developed by Araki~\cite{ArakiTT}, as shown in Fig.~\ref{fig:overint}.
We review Araki's argument in some detail in Appendix~\ref{app:araki}. Here we just mention some key elements.

The fact that $ \sX_{R_-} \lor \sX_{R_+}$ turns out to be larger than naively expected 
is reminiscent of the ``time-like tube theorem'', and in fact Araki developed these techniques to argue for an extension of the Borchers' formulation of the theorem~\cite{BorchersTT}, that allows for more general cases.
 An important property used in the argument is that the spectrum of the theory lies in the forward lightcone. 
It turns out that this is sufficient for the closure of the algebra of operators in certain regions extended in time to contain operators within a {\it spatially} extended region surrounding the original region considered. In the case of $ \sX_{R_-} \lor \sX_{R_+}$,  we obtain operators supported in a larger region, $A$ (such that $\sX_{R_-} \lor \sX_{R_+} = \sY_{A}$), that includes additional points between two hyperbolas outside of $\hat R_- \cup \hat R_+$. See figure~\ref{fig:overint}. Crucially, this is true without the need of any equation of motion.  We also note that this example does not follow from a direct application of Borchers' time-like tube theorem.\footnote{Note that the ``timelike envelope'' of $\hat R_- \cup \hat R_+$ is just $\hat R_- \cup \hat R_+$ itself. Araki's argument for the additional region requires stronger assumptions than those needed for Borchers' time-like tube theorem but does allow for the treatment of more general temporally extended regions.}

This is in stark contrast to the situation in which we start with operators supported in some~(infinitesimally temporally thickened) {\it spatial} subregion in which case, without equations of motion, one cannot conclude that closure of the algebra of operators associated to the original region contains operators supported at any new spacetime points. 
Thus informally we may say that ``{extending} time extends space, but {extending} space does not extend time'' in the case of operator algebras in a theory with no equations of motion.

Despite the subtlety that $\sX_{R_-}\vee\sX_{R_+}$ contains operators supported in the entire region $A,$ which includes additional boundary points outside of $\hat R_- \cup \hat R_+,$ for distinct overlapping intervals (i.e. $0 < b < a$) we always have the proper inclusion $A \subsetneq \widehat{R_- \cup R_+}.$ Thus, from the equivalence $\sX_{R_-\cup R_+} = \sY_{\widehat{R_- \cup R_+}}$ in this case we clearly have superadditivity, $\sX_{R_-}\vee\sX_{R_+} (=\sY_A) \subsetneq \sX_{R_- \cup R_+} (= \sY_{\widehat{R_- \cup R_+}}).$

Now consider increasing the separation between the two intervals, $b$. For $b=a$, the intervals just touch each other, and the story remains the same, namely $\sX_{R_-}\vee \sX_{R_+} = \sY_{A} \subsetneq \sX_{R_- \cup R_+} = \sY_{\widehat{R_- \cup R_+}}$.\footnote{Here with $R_- = \{x^1 \in (-a, 0)\},$ $R_+ = \{x^1 \in (0,a)\}$, we define $R_-\cup R_+ = \{x^1 \in (-a, a)\}$
with the point $x^1=0$ included. That this point can be included comes from ``strong additivity'' of the $\sB^{(N)}_R$ algebras of the finite $N$ CFT. Namely,  $B^{(N)}_{R_-} \vee B^{(N)}_{R_+} = B^{(N)}_{R_- \cup R_+}$. $\sX_{R_+  \cup R_+}$ is defined as the 
large $N$ limit of $B^{(N)}_{R_- \cup R_+}$, also with $x^1=0$ point included.
} The only difference is that the additional points disappear, i.e. now $A = \hat R_- \cup \hat R_+$. 

For $b > a$, the two intervals no longer overlap, and there are some new elements. Geometrically, we now have 
 $\hat R_+ \cup \hat R_- = \widehat{R_+ \cup R_-}$, and as a result $\sX_{R_+} \lor \sX_{R_-} = \sY_{\hat R_+ \cup \hat R_-} = \sY_{\widehat{R_+ \cup R_-}}$.  
There are no extra operators in $\sX_{R_+} \lor \sX_{R_-}$ beyond those in $\hat R_+ \cup \hat R_-$. 
In this case, it is not possible to obtain $\sX_{R_+ \cup R_-}$ directly from the CFT, and we have to resort to holography. 

\begin{figure}[h]
\begin{centering}
\includegraphics[width=0.5\textwidth]{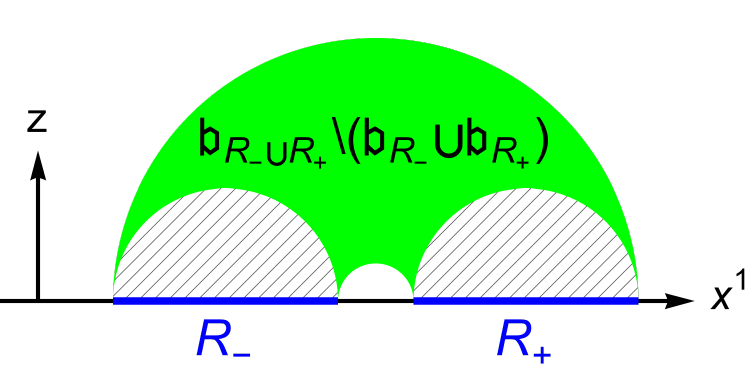}
\end{centering}
\caption[]{\small The $x^0=0$ slice of Poincar\'e AdS. Extra operators in $\sX_{R_- \cup R_+}$ that are not in $\sX_{R_-}\vee\sX_{R_+}$ are realized geometrically in the bulk as operators supported in the green shaded region.
}
\label{fig:entShadow}
\end{figure}

The end points of $R_\pm$ define a cross ratio
\be 
\eta = {a^2 \ov b^2}  \in (0,1) \ .
\ee
For $\eta > \ha$, the entanglement wedge for $R_+ \cup R_-$ is connected,~i.e. there is a proper inclusion $\fb_{R_+} \cup \fb_{R_-} \subsetneq 
\fb_{R_+\cup R_-}$. Thus there is superadditivity $\sX_{R_+} \lor \sX_{R_-} \subsetneq \sX_{R_+ \cup R_-}$. 
Here the superadditivity cannot be understood geometrically from the boundary theory, and it implies that there are nonlocally generated operators in the joint algebra of the two intervals. The bulk description geometrizes these additional operators, which are located in the region $\fb_{R_+ \cup R_-}\backslash \le(\fb_{R_+} \cup \fb_{R_-}\ri)$ (see the green shaded part in Fig.~\ref{fig:entShadow}). 
This region is also a Python's lunch~\cite{Brown:2019rox}, and operators in it should be reconstructible in the boundary theory using modular flow {for $\sX_{R_+ \cup R_-}$}~\cite{Faulkner:2017vdd}. 

When we increase $b$ further, eventually $\eta < \ha$, we have $\fb_{R_-} \cup \fb_{R_+} = \fb_{R_- \cup R_+}$ and $\sX_{R_-} \vee \sX_{R_+} = \sX_{R_- \cup R_+}$. There is no longer superadditivity.

\begin{figure}[h]
\begin{centering}
\includegraphics[width=0.3\textwidth]{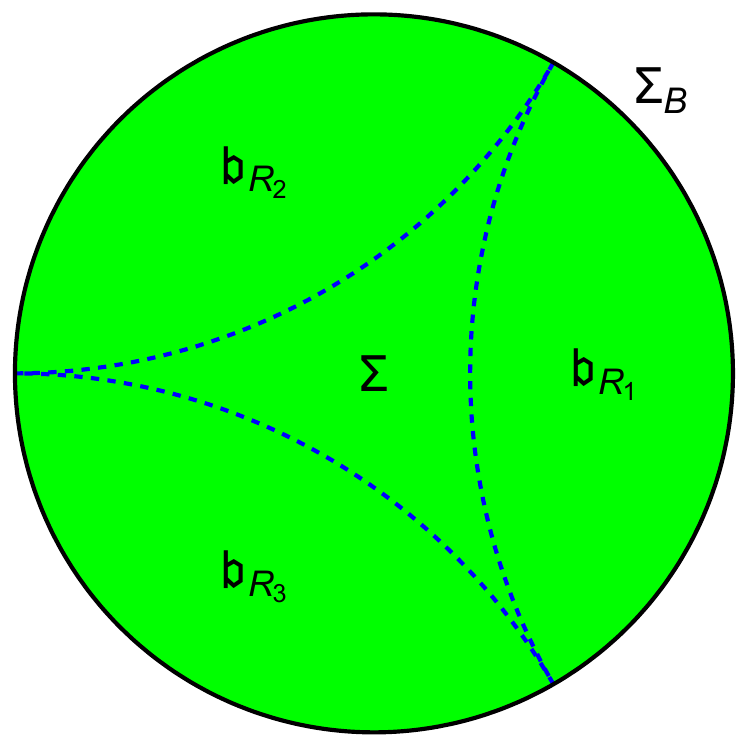}
\end{centering}
\caption[]{\small The superadditivity $\vee_{i=1}^{3} \sX_{R_i} \subsetneq \sX_{B}$ is geometrically realized in the bulk by $\cup_{i=1}^{3} \fb_{R_i}$ (the union of the regions between the dashed blue lines and the boundary) being a proper subset of the full bulk Cauchy slice $\Sigma$ (the entire green disk).
}
\label{fig:notAllBulk}
\end{figure}

The discussion can be generalized to multiple intervals. For example, consider separating a boundary Cauchy slice $\Sig_B$ into 
multiple intervals 
\be 
\Sig_B = \cup_i R_i \ .
\ee
Let $\sX_B$ denote the large $N$ limit of the algebra associated with $\Sig_B$ (i.e. large $N$ limit of the full boundary algebra).  We then have  
 \be 
 \lor_i \sX_{R_i} \subsetneq \sX_B  ,
 \ee
which in the bulk corresponds to 
\be 
\cup_i \fb_{R_i} \subsetneq \Sig  \ .
\ee
Here $\Sig$ is the full bulk Cauchy slice ending on the boundary at $\Sig_B$, and $\fb_{R_i}$ is the entanglement wedge for $R_i$. See Fig.~\ref{fig:notAllBulk}.

\subsubsection{Overlapping spherical regions in the vacuum of CFT$_d$}

The example of Fig.~\ref{fig:overint} can be generalized to higher dimensions, $d>2,$ with some interesting differences. 
Consider two ball-shaped regions, $R_{\pm},$ of radius $a > 0$ centered at coordinates $x^1 = \pm b, \, x^j = 0,\, j = 2, ..., d-1$ on the $x^0=0$ time-slice of $d-$dimensional Minkowski space. For $b < a$ these regions overlap. The causal completion of each ball is a double-cone
\be \label{eq:defDoubleConesPM}
	\hat{R}_{\pm} = \le\{(x^0,x^1,x^j) ~\biggr|~ |x^0| + \sqrt{(x^1 \mp b)^2 + \sum_{j=2}^{d-1} (x^j)^2} < a \ri\} \ .
\ee

To each of these balls a von Neumann algebra, $\sX_{R_\pm}$ is assigned.
In the vacuum state, we have $\sX_{R_{\pm}} = \sY_{\hat R_{\pm}}$, i.e. both algebras are generated by single-trace operators localized in the corresponding double-cone. Araki's argument~\cite{ArakiTT} applied to this case shows that  $\sX_{R_+} \lor \sX_{R_-}$ is given by the algebra generated by single-trace operators localized in the causal completion of $R_+ \cup R_-$~(see Appendix~\ref{app:araki} for details), i.e. 
\be 
\sX_{R_+} \lor \sX_{R_-} = \sY_{\hat R_+} \lor \sY_{\hat R_-} = \sY_{\widehat{R_+ \cup R_-}} \ .
\ee
Note that {we again have} $\hat R_+ \cup \hat R_- \subsetneq  \widehat{R_+ \cup R_-} = (\hat R_+ \cup \hat R_-)''$; however, the region $A$ such that $\sY_{\hat R_- \cup \hat R_+} = \sY_A$ now coincides with $\widehat{R_- \cup R_+}.$ 
That is, the minimal completion of $\sY_{\hat R_+} \cup \sY_{\hat R_-}$ into a von Neumann algebra is now the same as causal completion. 
For $d > 2$, $R_+ \cup R_-$ is not a spherical region so we expect 
\be 
 \sY_{\widehat{R_+ \cup R_-}}  \subsetneq \sX_{R_+ \cup R_-} \ .
\ee
Therefore we expect
\be 
\sX_{R_+} \lor \sX_{R_-} \subsetneq \sX_{R_+ \cup R_-} \ .
\ee

While we again have superadditivity, the way it is realized is different from $d=2$. Recall that for $d=2$ we found
$\sX_{R_+} \lor \sX_{R_-} \subsetneq  \sY_{\widehat{R_+ \cup R_-}}  =\sX_{R_+ \cup R_-} $, while for $d >2$ we have $\sX_{R_+} \lor \sX_{R_-} = \sY_{\widehat{R_+ \cup R_-}} \subsetneq \sX_{R_+ \cup R_-}$, and the superadditivity cannot be seen geometrically on the boundary. 

The discussion for $b > a$, i.e. disjoint spherical regions, is similar to that of $d=2$. There is superadditivity until $b$ is larger than some critical value.

\subsubsection{Overlapping regions at finite temperature for $d=2$} \label{sec:finTDeq2}

We now consider the overlapping intervals of Fig.~\ref{fig:overint} in a CFT$_2$ at a finite temperature.\footnote{Recall that $R_\pm$ are respectively centered at $x^1=\pm b$ with half-width $a>b$.}
 In this case it is not possible to use Araki's argument to obtain $\sX_{R_+} \lor \sX_{R_-}$, as it assumes that the spectrum lies in the forward lightcone, which is not satisfied at finite temperature. Nevertheless we can use the help of holography to work out $\sX_{R_+} \lor \sX_{R_-}$ explicitly, and see that superadditivity occurs.

The dual bulk geometry is that of the BTZ black hole\footnote{Since we consider the boundary spatial direction to be noncompact, the bulk geometry in fact coincides with that of an AdS-Rindler region.}
\bea 
	ds^2 &=&  -f(r)dt^2 + f(r)^{-1} dr^2 + {r^2 \ov L_{AdS}^2} dx^2 \\
	&=& {L_{\rm AdS}^2 \ov w^2} \le(-\le(1-w^2\ri) \le({2\pi \ov \beta}\ri)^2 dt^2 + {dw^2 \ov 1-w^2} + \le({2\pi \ov \beta}\ri)^2 dx^2\ri) \ ,
\eea
where $f(r) = (r^2 - r_+^2)L_{\rm AdS}^{-2}$, with $r_+$ the location of the horizon, and $r \to \infty$ is the boundary~(i.e. $r \in (r_+, \infty)$)~\cite{Banados:1992wn}. The inverse temperature (with respect to $t$) is $\beta = 2\pi L_{AdS}^2/r_+$.  We also defined a dimensionless radial coordinate $w = r_+/r \in (0,1)$ with $w=0$ the asymptotic boundary and $w=1$ the horizon.

\begin{figure}[h]
        \centering
        \begin{subfigure}[b]{0.45\textwidth}
            \centering
		\includegraphics[width=\textwidth]{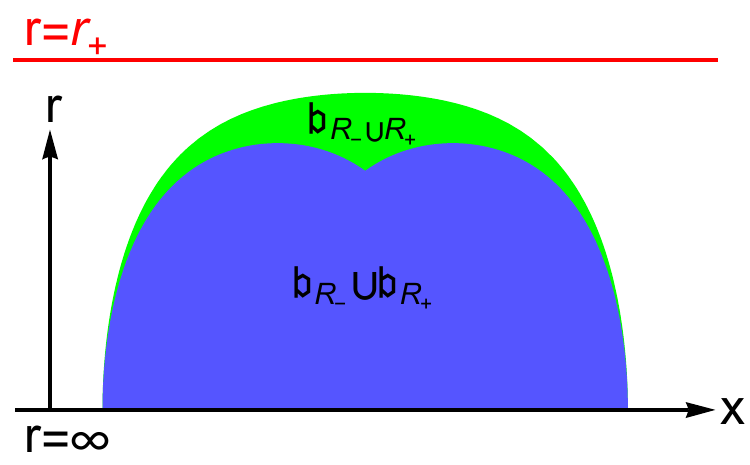} 
            \caption[]%
            {{\small $t=0$ slices of the bulk duals of $\sX_{R_-}\vee\sX_{R_+}$ and $\sX_{R_- \cup R_+}$.} }    
        \end{subfigure}
        \hfill
        \begin{subfigure}[b]{0.45\textwidth}   
            \centering 
		\includegraphics[width=\textwidth]{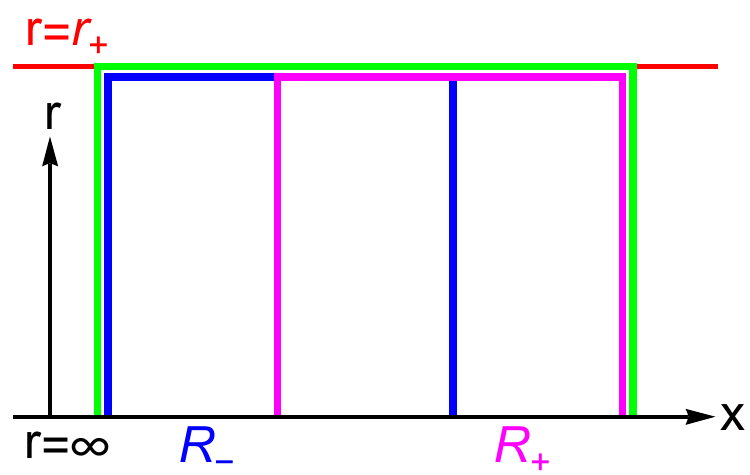}
            \caption[]%
            {{\small RT surface $R_-$ (blue), $R_+$ (magenta), and $R_- \cup R_+$ (green) as $\beta \to 0$.}}    
        \end{subfigure}
        \caption[  ]
        {\small  Bulk subregions on the $t=0$ slice of the dual at finite temperature. The duals of $\sX_{R_-} \vee \sX_{R_+}$ and $\sX_{R_-\cup R_+}$ are, respectively, the blue and the union of blue and green shaded regions in (a). The mutual information is computed holographically as the sum of areas of the blue and magenta surfaces minus that of the green surface in (b). In the high temperature limit, this is equal to the area of the RT surface for the region $R_+ \cap R_-.$} 
\label{fig:blackBraneBulk}
\end{figure}

The $\sX_{R_\pm}$ are respectively equal to the bulk algebras of the entanglement wedges of $R_\pm$. 
Since the bulk algebras are additive,  $\sX_{R_+} \lor \sX_{R_-}$ is given by the bulk algebra in the blue shaded region of Fig.~\ref{fig:blackBraneBulk}(a), which is $\fb_{R_+} \cup \fb_{R_-}$, i.e.  $\sX_{R_+} \lor \sX_{R_-} = \widetilde \sM_{\fb_{R_+} \cup \fb_{R_-}}$. 

\begin{figure}[h]
\begin{centering}
	\includegraphics[width=2.5in]{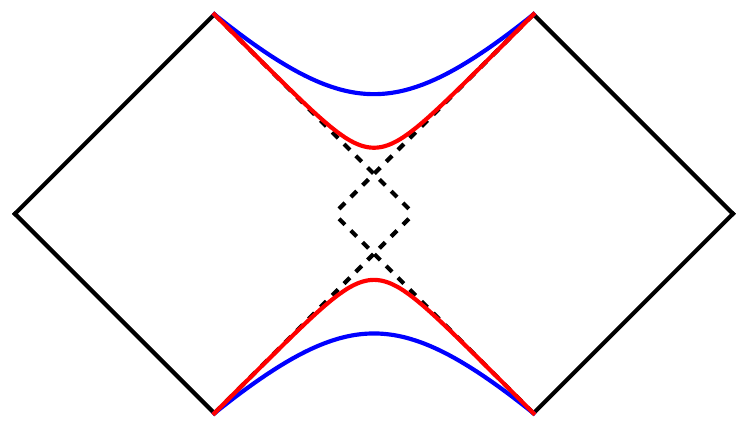}
\par\end{centering}
\caption{The boundary region $A$ 
whose associated algebra, $\sY_A,$ is equal to $\sX_{R_+} \vee \sX_{R_-}$. The vacuum 
case (bounded by the solid black lines and blue curves) was given in Fig.~\ref{fig:overint} and is shown here for comparison. 
At a finite temperature, $A$ is the region bounded by the solid black lines and red curves. The dashed lines indicate the boundaries of the original diamonds $\hat R_{\mp}$. The figure is drawn for $a = \frac{5}{\pi}\beta,~ b = \frac{4}{\pi} \beta.$
}
\label{fig:fiT}
\end{figure}

From the discussion of Sec.~\ref{sec:qec}, 
$\widetilde \sM_{\fb_{R_+} \cup \fb_{R_-}}$ is equal to $\sY_A,$ where $A$ is a boundary subregion whose causal domain is $\widehat{\fb_{R_+} \cup \fb_{R_-}}.$ $A$ can be obtained by sending light rays from $\fb_{R_+} \cup \fb_{R_-}$ to the boundary.
We leave the details to Appendix~\ref{app:araki}. The result is shown in Fig.~\ref{fig:fiT}.

Interestingly, the region $A$ obtained at finite temperature is smaller than that obtained in the vacuum. Thus superadditivity becomes more ``pronounced'' at finite temperature compared to the vacuum state. The larger the temperature, the smaller $A$ is.
In the low temperature limit, $A$ approaches the vacuum result (i.e. in figure~\ref{fig:fiT} the red curves approach the blue ones), while in the infinite temperature limit 
$A \to \hat R_+ \cup \hat R_-$ (i.e. in figure~\ref{fig:fiT} the red curves approach the dashed black lines).

The temperature dependence of $A$ suggests that we may interpret the equality
$\sX_{R_+} \lor \sX_{R_-} =\sY_A,$ where $A$ is strictly larger than  $\hat R_+ \cup \hat R_-,$ as reflecting ``quantum correlations'' between $R_+$ and $R_-$.  Increasing the temperature tends to ``destroy'' quantum correlations, 
and thus the fact that $A$ is closer to $\hat R_+ \cup \hat R_-$ at higher temperatures, suggests that the proper containment $\hat R_+ \cup \hat R_- \subsetneq A$ is a reflection of quantum correlations. 
To see this more explicitly, recall that, at finite temperature, connected equal-time two-point functions decay exponentially with the distance $L$ for $L \gg \beta$. As $\beta \to 0$,  correlation functions for any finite $L$ essentially vanish and precisely in this limit $A$ no longer contains extra points, since $A \to \hat R_- \cup \hat R_+$. 

Putting in a boundary UV cutoff, we can also see the loss of quantum correlations using the mutual information $I(R_+ : R_-)$ between $R_+$ and $R_-$, 
 \be
	I(R_+ : R_-) = S(R_+) + S(R_-) - S(R_+ \cup R_-)  ,
\ee
where $S(R_+)$ is the entanglement entropy of $R_+$. $I (R_+ : R_-)$ can be readily calculated 
using holographic entanglement entropy, and is given {in the $\b \to 0$ limit} by~\cite{Hubeny:2013gta} 
\be
I (R_+ : R_-) =
S(R_+ \cap R_-) + O\le(e^{-{1\ov\beta}}\ri), 
\ee
i.e. the entropy of the overlap region. See Fig.~\ref{fig:blackBraneBulk} (b).
 This is exactly the result that would be obtained for the CFT being in a factorized state on the regions $R_+\setminus(R_+\cap R_-),~R_+\cap R_-,$ and $R_-\setminus(R_+\cap R_-)$, i.e. if the state had the form $\rho_{R_+\cup R_-} = \rho_{R_+\setminus(R_+\cap R_-)} \otimes \rho_{R_+ \cap R_-} \otimes \rho_{R_-\setminus(R_+\cap R_-)},$ so that there are no correlations between separate subregions. In this high-temperature limit where the state factorizes the additive algebra does not contain any boundary operators at extra spacetime points.

\subsubsection{Probing the interior of a black hole using superadditivity} \label{sec:insideHor}

We now discuss an interesting case where superadditivity can be used to probe the interior of a black hole. 
Consider the thermofield double state on two copies of $\mathbb{R} \times S^1.$ Above the Hawking-Page temperature, the bulk dual is the BTZ black hole~\cite{Banados:1992wn}. Denote the entire $t=0$ Cauchy slice of the `left' CFT by $L,$ and an angular interval of half-width $w$ on the $t=0$ slice of the `right' CFT by $R_w.$ 

\begin{figure}[h]
\begin{centering}
\includegraphics[width=0.35\textwidth]{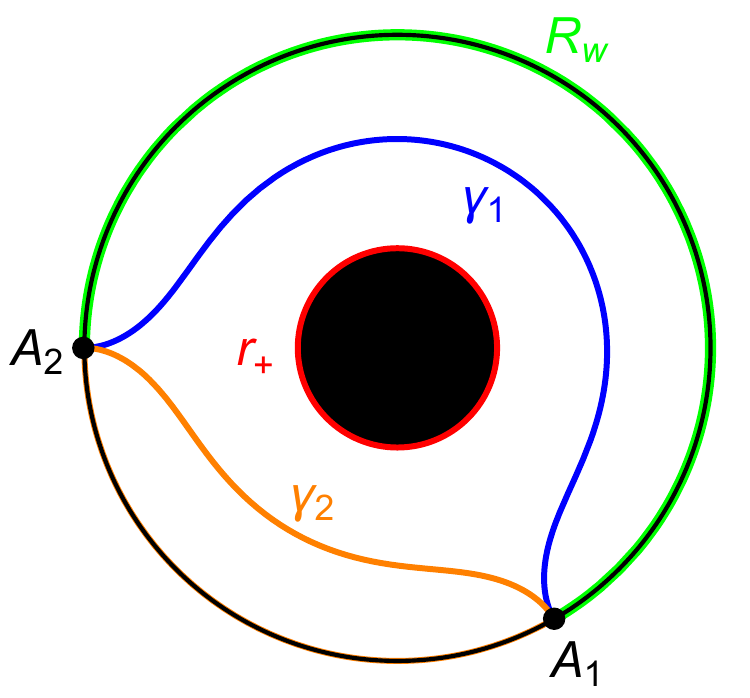}
\end{centering}
\caption[] {\small $t=0$ slice of the right exterior of the BTZ black hole. For the boundary interval $R_w$ of half-width $w,$ there are two competing extremal surfaces, $\ga_1$ and $\ga_2 \cup r_+,$ homologous to $R_w.$ For $w < w_c,$ $\ga_1$ is the RT surface while for $w > w_c,$ $\ga_2 \cup r_+$ is the RT surface.} 
\label{fig:explainWc}
\end{figure}

The entanglement wedge for $R_w$ undergoes a transition at a critical value $w_c$~\cite{Ryu:2006bv}. 
More explicitly, suppose the end points of $R_w$ are labelled as $A_1$ and $A_2$. There are two extremal surfaces (curves) connecting $A_1$ and $A_2$, which we denote as $\ga_1$ and $\ga_{2}$ (see Fig.~\ref{fig:explainWc}), with corresponding lengths $\ell_1$ and $\ell_2$. $w_c$ is defined as the value of the boundary half-width for which $\ell_1 = 2 \pi r_+ + \ell_2$, where $r_+$ is radius of the horizon. For $w < w_c$, we have 
\be 
\ell_{1} <  2 \pi r_+  + \ell_{2}  ,
\ee
the RT surface for $R_w$ is given by $\ga_1$ and its entanglement wedge and causal domain coincide. 
For $w > w_c$, we have 
\be 
\ell_{1} >  2 \pi r_+  + \ell_{2}  ,
\ee
the RT surface for $R_w$ is given by the union of $\ga_2$ and the horizon, and 
the entanglement wedge of $R_w$ exceeds its causal domain. See Fig.~\ref{fig:explainWc}.
The explicit expression for $w_c$ is given by~\cite{Hubeny:2013gta}
\be 
	w_c = {\beta \ov 2\pi l} \coth^{-1} \le(2\coth\le({2\pi^2 l \ov \beta}\ri)-1\ri) \ ,
\ee 
where $l$ is the radius of the boundary circle. Note that in the high temperature limit, 
$w_c \to \pi$.  
Similarly, at $w= \pi - w_c$, the entanglement wedge of the complement of $R_w,$ $\bar R_w$ (an angular interval of half-width $\pi-w$), undergoes a transition. 

\begin{figure}[h]
  \centering
        \begin{subfigure}[b]{0.3\textwidth}
            \centering
			\includegraphics[width=\textwidth]{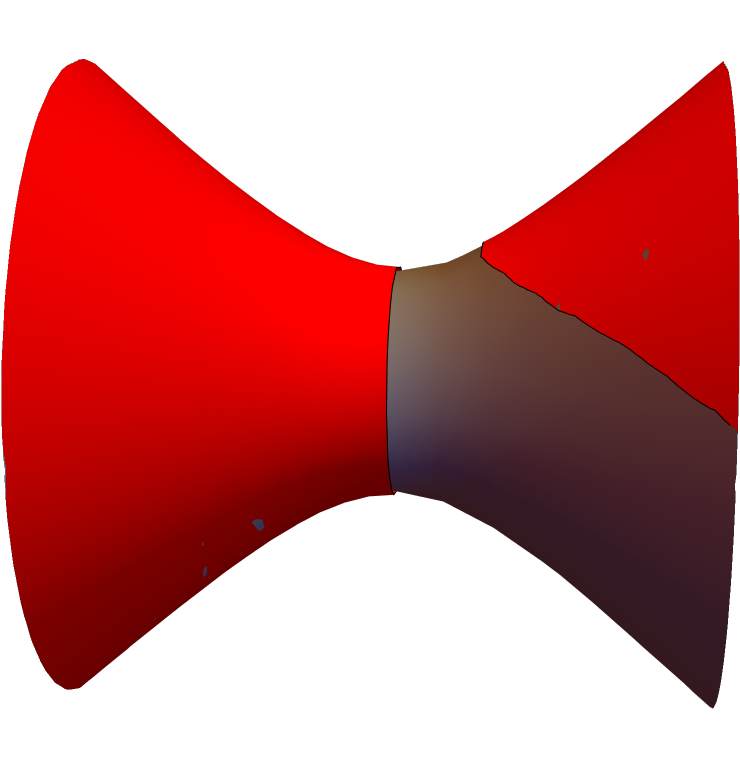} 
            \caption[]%
            {{\small Bulk dual of $\sX_L \vee \sX_{R_{w}}$, whose domain of dependence lies outside the horizon.}}    
        \end{subfigure}
       \qquad\qquad\qquad\qquad
        \begin{subfigure}[b]{0.3\textwidth}   
            \centering 
			\includegraphics[width=\textwidth]{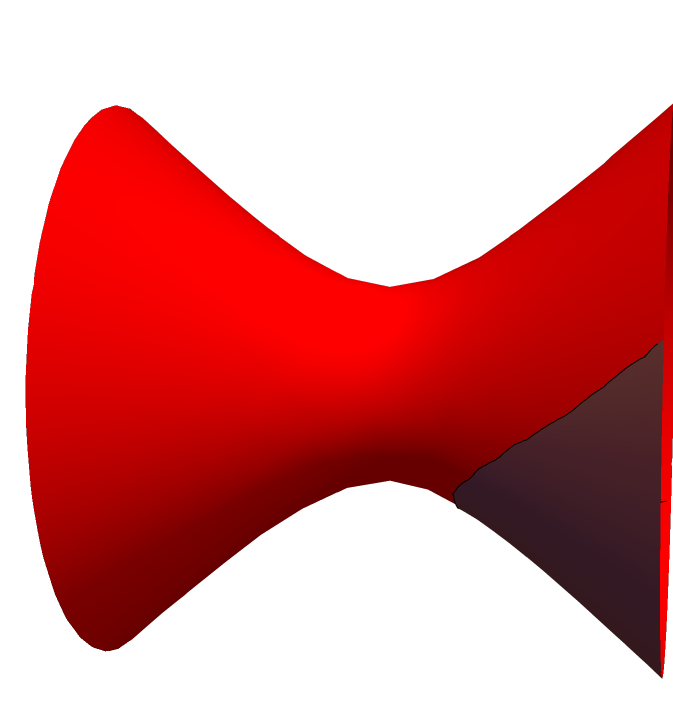}
            \caption[]%
            {{\small Bulk dual of $\sX_{L\cup R_w}$, whose domain of dependence includes a region behind the horizon.}}    
        \end{subfigure}
        \caption[  ]
        {\small $t=0$ slices of the bulk duals of (a) $\sX_L \vee \sX_{R_{w}}$ and (b) $\sX_{L\cup R_w}$ are shown in red.
        }  
\label{fig:bulkDualAddBTZ}
\end{figure}

\begin{figure}[h]
  \centering
        \begin{subfigure}[b]{0.45\textwidth}
            \centering
			\includegraphics[width=\textwidth]{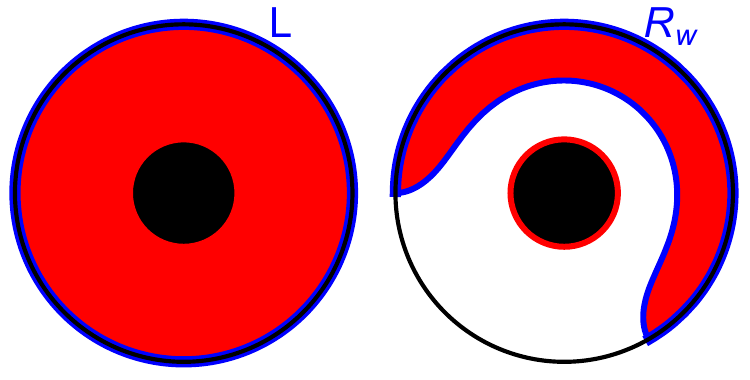} 
            \caption[]%
            {{\small Bulk dual of $\sX_L \vee \sX_{R_{w}}$.}}    
        \end{subfigure}
        \hfill
        \begin{subfigure}[b]{0.45\textwidth}   
            \centering 
			\includegraphics[width=\textwidth]{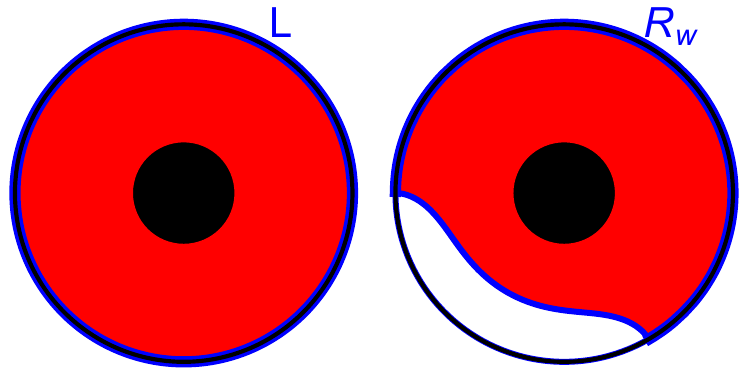}
            \caption[]%
            {{\small Bulk dual of $\sX_{L\cup R_w}$.}}    
        \end{subfigure}
        \caption[  ]
        {\small $t=0$ slices of the bulk duals of (a) $\sX_L \vee \sX_{R_{w}}$ and (b) $\sX_{L\cup R_w}$ are shown in red.
        }  
\label{fig:flatBulkDualAddBTZ}
\end{figure}

\begin{figure}[h]
\begin{centering}
\includegraphics[width=0.4\textwidth]{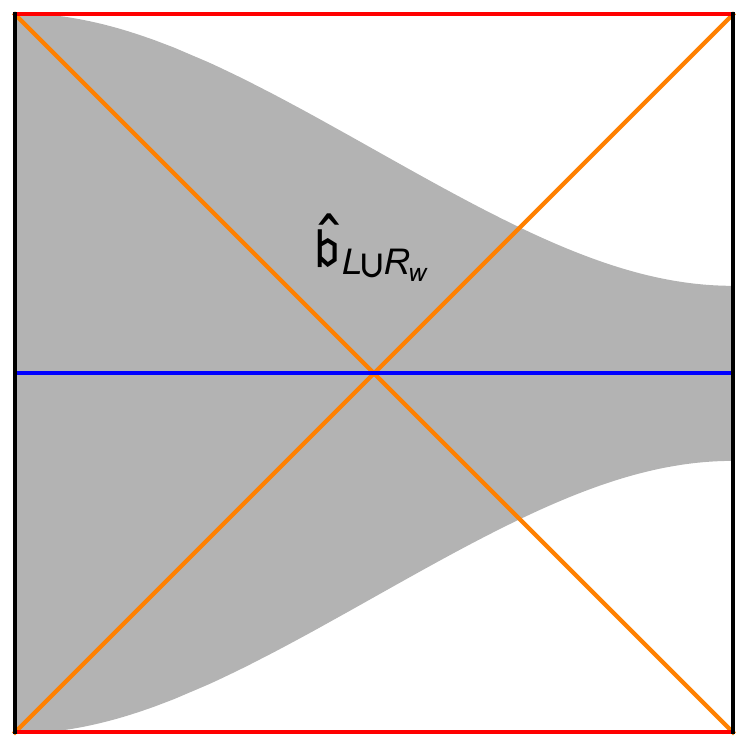}
\end{centering}
\caption[] {\small $\theta = 0$ slice of the bulk dual of $\sX_{L \cup R_w}$ drawn on the Penrose diagram of the BTZ black hole. We explicitly see that points behind the horizon are included.} 
\label{fig:seeInsideSlice}
\end{figure}

Now consider $w$ in the range $\pi - w_c < w < w_c$, for which the RT surface for $R_w$ is given by $\ga_1$ (from $w < w_c$). The RT surface for $L \cup R_w$ is then given by $\ga_2$, as $\ell_2 < \ell_1 + 2 \pi r_+$~(from $\pi-w_c < w$), with the corresponding entanglement wedge $\fb_{L \cup R_w}$ indicated in Figs.~\ref{fig:bulkDualAddBTZ} and~\ref{fig:flatBulkDualAddBTZ} (b). In contrast, the bulk dual of $\sX_{L} \vee \sX_{R_w},$ $\fb_L \cup \fb_{R_w},$ is shown in Figs.~\ref{fig:bulkDualAddBTZ} and~\ref{fig:flatBulkDualAddBTZ} (a). We thus have $\fb_{L} \cup \fb_{R_w} \subsetneq \fb_{L \cup R_w}$, which implies 
superadditivity 
\be \label{eiu}
\sX_L \lor \sX_{R_w} \subsetneq \sX_{L \cup R_w} \ .
\ee
The additional operators are located in the portions of the red bulk regions shown in Figs.~\ref{fig:bulkDualAddBTZ} (b) and~\ref{fig:flatBulkDualAddBTZ} (b) that are not red in Figs.~\ref{fig:bulkDualAddBTZ} (a) and~\ref{fig:flatBulkDualAddBTZ} (a). 
From additivity of the bulk algebras, $\sX_L \lor \sX_{R_w}$ describes operators in the bulk region $\hat \fb_L \cup \hat \fb_{R_w}$, which lies completely outside the horizon. 
In contrast, $\widehat{ \fb_{L \cup R_w}}$  includes a region behind the horizon, as indicated in Fig.~\ref{fig:seeInsideSlice}. 
Thus, in this setting, superadditivity allows us to probe the interior of a black hole. 

When $w > w_c$, the RT surface for $R_w$ is given by the union of $\ga_2$ and the horizon, and that of $R_w \cup L$ is given by $\ga_2$. We then have $\fb_{R_w} \cup \fb_L = \fb_{R_w \cup L}$, both $\sX_L \lor \sX_{R_w}$ and $\sX_{L\cup R_w}$ probe the black hole interior, and there is no superadditivity. 
For $w < \pi - w_c$,  the RT surface for $R_w$ is given by $\ga_1$, while that of $R_w \cup L$ is given by 
the union of $\ga_1$ and the horizon (as $\ell_1 + 2 \pi r_+ < \ell_2$). Now neither $\sX_L \lor \sX_{R_w}$ nor $\sX_{L\cup R_w}$ probes the black hole interior and again there is no superadditivity.

One interesting feature of this example is that superadditivity occurs for any choice of half-width $w \in (0, \pi)$ in the high-temperature ($\beta \to 0$) limit since $w_c \to \pi$. In particular, with access to the full `left' CFT and only a proper subregion (no matter how large) of the `right' CFT one can never probe the black hole interior with the additive algebra but always can probe the interior with the joint algebra. Thus at high-temperature superadditivity is ``more pronounced.''
This is consistent with the observed `enhancement' of superadditivity at finite temperature for overlapping intervals discussed in section~\ref{sec:finTDeq2}. 

In contrast, note that the low temperature limit is no longer described by the BTZ geometry. Below the Hawking-Page temperature the bulk is described by two copies of thermal AdS, in which case the Hilbert space factorizes and $\sX_L$ and $\sX_{R_w}$ act on different Hilbert space factors.  Superadditivity for $L \cup R_w$ never occurs.

To summarize, above the Hawking-Page temperature one finds that the boundary subregion $L\cup R_w$ never probes the black hole interior for $w < \pi - w_c$ and always probes the black hole interior for $w > w_c$. In the intermediate regime $\pi- w_c < w < w_c$ only the joint algebra $\sX_{L \cup R_w}$ probes the interior while the additive algebra $\sX_L \vee \sX_{R_w}$ does not, providing a striking example of the superadditvity phenomenon. Interestingly, since $w_c > \pi/2$, the additive algebra always needs more than half of the `right' boundary to probe behind the horizon while the superadditive can probe the interior with less than half of the `right' boundary~(and even with an arbitrarily small piece in the high-temperature limit).

\section{Implications of superadditivity (I): quantum error correcting properties} \label{sec:impQEC}

In this and the following section we discuss physical implications of the superadditivity property~\eqref{hen3} in holographic large $N$ field theories. First, in this section, we show that superadditivity is the physical origin of the quantum error correcting properties previously postulated for holographic systems, and comment on tensor network and abstract algebraic models based on quantum error correcting codes.

\subsection{Explanation of quantum error correcting properties from superadditivity}

\begin{figure}[h]
  \centering
        \begin{subfigure}[b]{0.45\textwidth}
            \centering
			\includegraphics[width=\textwidth]{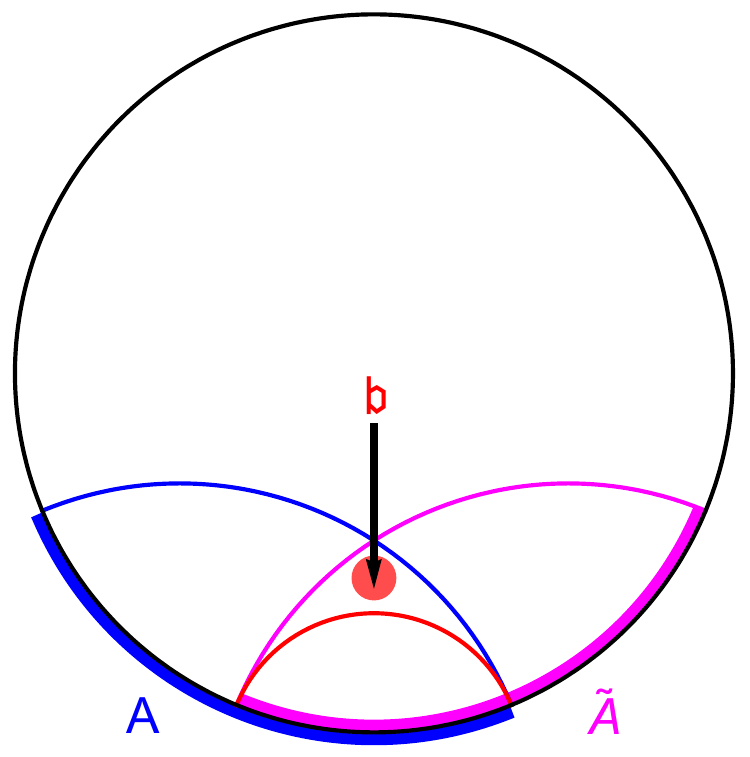} 
            \caption[]%
            {{\small $\sX_A \cap \sX_{\tilde A} \supsetneq \sX_{A \cap \tilde A} \Leftrightarrow \fb_A \cap \fb_{\tilde A} \supsetneq \fb_{A \cap \tilde A}$}}    
        \end{subfigure}
        \hfill
        \begin{subfigure}[b]{0.45\textwidth}   
            \centering 
			\includegraphics[width=\textwidth]{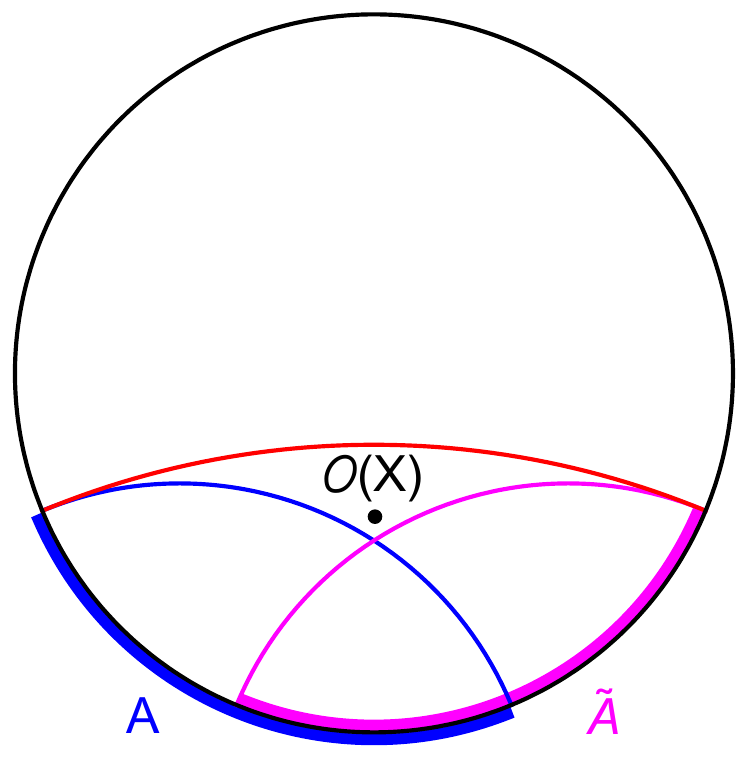}
            \caption[]%
            {{\small  $\sX_A \vee \sX_{\tilde A} \subsetneq \sX_{A \cup \tilde A} \Leftrightarrow \fb_A \cup \fb_{\tilde A} \subsetneq \fb_{A \cup \tilde A}$}}    
        \end{subfigure}
        \caption[  ]
        {\small Here the figure should be understood as representing a single time slice in AdS with the black circle denoting the boundary while the interior of the circle is the bulk of AdS.
(a) Quantum operations in a local subregion $\fb$ are encoded in the boundary system in a highly redundant way as 
they can be contained in the entanglement wedges of many possible boundary subregions, such as $A$ and $\tilde A$, 
and in principle infinitely many others. Furthermore, $\fb$ can lie in the intersection of entanglement wedges of $A$ and $\tilde A$, but not in the entanglement wedge of $A \cap \tilde A$. 
  (b) That $\fb_A \cup \fb_{\tilde A} \subsetneq \fb_{A\cup \tilde A}$ implies that there exist operations, e.g. the operator $O(X)$, that cannot be reconstructed from those in either $A$ or $\tilde A$, but can be reconstructed from those $A \cup \tilde A$.
        }  
\label{fig:redu}
\end{figure}

An implication of entanglement wedge reconstruction is that there are many redundant boundary descriptions of any bulk operator, see Fig.~\ref{fig:redu}. Such redundant ``encodings'' have been interpreted ``phenomenologically'' in terms of quantum error correction~\cite{Almheiri:2014lwa}.\footnote{See~\cite{Verlinde:2012cy} for an earlier discussion of using quantum error correction to describe black hole states.} 
Heuristically, the ability to recover all the quantum information in a bulk subregion, $\fa$, with knowledge of only the subregion $A$ on the boundary implies that the information in $\fa$ is robust even if we ``make errors'' in or completely ``erase''  the complement, $\bar A,$ of $A$.
Thus any quantum information in a local bulk region is stored on the boundary in a highly redundant way. Furthermore, equation~\eqref{Aoo2} implies that the union of two subsystems $R_1$ and $R_2$ contains more {operations} than the minimal ``combination'' of the respective {operations} for two subsystems.

The quantum error correction perspective has provided 
an important guiding principle to construct toy models of holographic duality using systems of finite dimensional Hilbert spaces, such as various types of tensor networks~\cite{Pastawski:2015qua, Hayden:2016cfa}. 
It has also inspired algebraic descriptions of holography using abstract quantum error correcting codes~\cite{Harlow:2016vwg, Kang:2018xqy, Kang:2019dfi, Gesteau:2020hoz, Gesteau:2020rtg, Gesteau:2021jzp, Faulkner:2020hzi}. 
However, it has not been clear why a semi-classical state of the boundary CFT gives rise to an error correcting code, or how to derive its error correcting properties.

Subregion-subalgebra duality and the superadditivity~\eqref{hen3} of the algebras $\sX_{R}$ 
provide a boundary derivation/explanation of the quantum error correcting properties indicated in Fig.~\ref{fig:redu}, and give a more precise formulation of them.
The redundancies in the boundary description simply come from the fact that a local bulk operator belongs to many different bulk subregions and thus can be represented using many different boundary subalgebras. 
The superadditivity of entanglement wedges indicated in Fig.~\ref{fig:redu} is a bulk geometric reflection of
superadditivity of the corresponding boundary algebras---the union of two boundary subsystems $R_1$ and $R_2$ contains more operations than the minimal ``combination'' of the respective operations for two subsystems.

In the remainder of this section, we analyze the quantum error correction (QEC) code-based approach to holography using the perspectives of subregion-subalgebra duality and superadditivity. As we will review in Sec.~\ref{sec:QEC}, the QEC approach is sensibly formulated only at the level of a finite but large $N$, with $N$ treated non-perturbatively. However, this formulation can only be approximate, and its approximate nature is currently not precisely controlled. The main difficulties stem from extending entanglement wedge reconstruction---originally formulated perturbatively in the $1/N$ ($G_N$) expansion---to a finite $N$ (finite $1/G_N$).

We therefore first comment on the difficulties in extending and the necessarily approximate nature of entanglement wedge reconstruction at finite $N$.
We then make comments on various quantum error correction code-based approaches to holography.

 \subsection{Entanglement wedge reconstruction at finite $N$} \label{sec:EWRFinN}

In this subsection, we discuss a possible formulation of entanglement wedge reconstruction at finite $N$.

\subsubsection{Approximate definition of the code subspace} \label{sec:approxCodeSpace}

We begin by discussing how to generalize the equivalence between the bulk Fock space and the boundary GNS Hilbert space to finite $N$. For definiteness, we will consider the system in the vacuum state $\ket{\Om}$, and have in mind the $\sN =4$ super-Yang-Mills theory as a prototypical example.

For a holographic system, in $1/N$ (or $G_N$) perturbation theory, the state space separates into disjoint sectors, each corresponding to a different classical solution. Each sector is represented by a bulk Fock space around a classical geometry, which is {\it equivalent} to the boundary GNS Hilbert space constructed around the state corresponding to that geometry, 
\be \label{ieu}
\sH_\Om^{\rm (Fock)}  =  \sH_\Om^{\rm (GNS)} \ .
\ee
By definition, the Fock space (or GNS Hilbert space) captures all low-energy physics around a specific geometry, with no overlap between Fock spaces of different geometries. 

The operator algebra acting on~\eqref{ieu} can be written as 
 \be \label{eb1}
\sB (\sH_\Om^{\rm (Fock)})  =  \sB (\sH_\Om^{\rm (GNS)})  =\pi_\Om( \sY )
 \ee
where $\sY$ is the algebra generated by single-trace operators, and $\pi_\Om$ denotes its representation on $\sH_\Om^{\rm (GNS)}$. The first equality of~\eqref{eb1} can be written explicitly in terms of the so-called global reconstruction, which expresses a bulk field operator $\phi(X)$, with $X$ denoting a point in the bulk spacetime, 
in terms of boundary operators as~\cite{Banks:1998dd,Hamilton:2006az} 
\bega \label{eun0}
\phi (X) =  \int d^{d} x \, K \le(X; x\ri)  \pi_{\Om}\le(\sO (x)\ri) = \pi_\Om (\sO_{G} (X)) \ ,\\
\sO_{G} (X) \equiv \int d^{d} x \, K \le(X; x\ri)  \sO (x), 
\end{gather} 
where $\sO (x)$ is the single-trace operator dual to $\phi$, $x$ denotes a boundary point, and $K (X;x)$ is the global reconstruction kernel.

\begin{figure}[!h]
\begin{center}
\includegraphics[width=5cm]{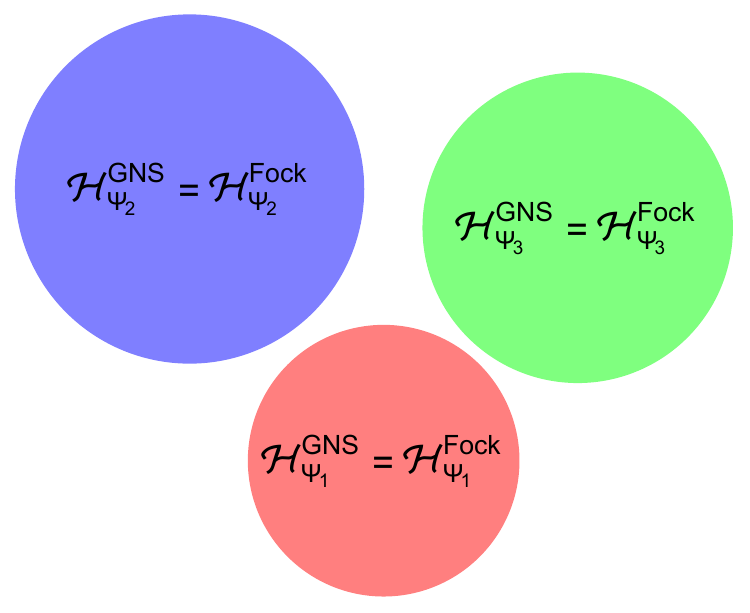}  \qquad\qquad\qquad
\includegraphics[width=5cm]{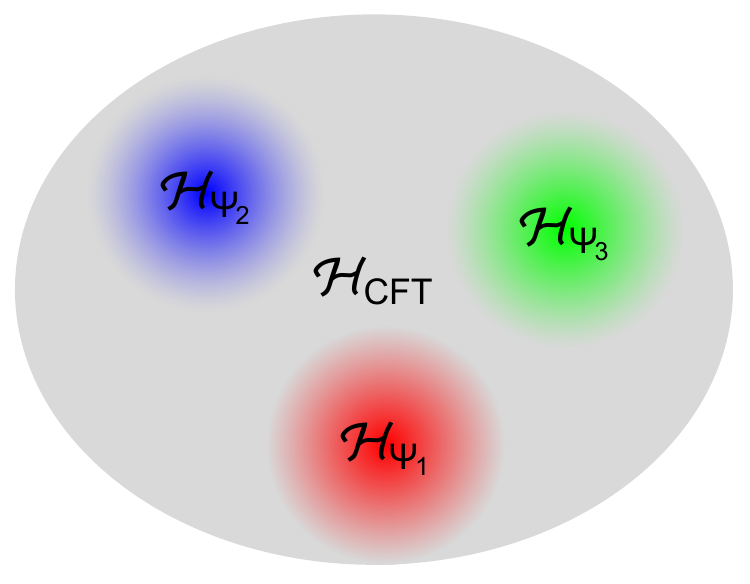}
\caption{Contrast of the structure of the space of states between infinite and finite $N$. Left: At infinite $N$, the state space separates into disjoint sectors. Right: At finite $N$, different sectors become connected and join to form a single $\sH_{\rm CFT}$. 
}
\label{fig:strH}
\end{center}
\end{figure}

Now consider a finite but large $N$, with $N$ treated {\it non-perturbatively.}  Heuristically, since bulk spacetime fluctuations are small, the semi-classical picture should remain approximately valid. We can still consider the space of small excitations around a classical geometry, but now different semi-classical sectors are no longer disconnected. Instead, they join to form a single Hilbert space $\sH_{\rm CFT}$, as illustrated in Fig.~\ref{fig:strH}. In this case, it is nevertheless convenient to isolate a subspace in the full Hilbert space $\sH_{\rm CFT}$ that describes the space of low-energy excitations around a bulk spacetime. 
To make connections with quantum error correction based approaches, we will refer to such a subspace as a code subspace.

We denote the space of excitations around the vacuum state $\ket{\Om}$ that captures  
low-energy physics around empty AdS as $\sH^{\rm (code)}_\Om \subset \sH_{\rm CFT}$. Specifying $\sH^{\rm (code)}_\Om$ precisely is challenging due to the need to work non-perturbatively in $G_N$. From the boundary perspective, states in $\sH^{\rm (code)}_\Om$ are, by definition, exact quantum states at a finite $N$, and thus we we do not have analytic control over them. 

An approximate description of $\sH^{\rm (code)}_\Om$ can be given as follows. Take the Fock space $\sH_\Om^{\rm (Fock)}$ around empty AdS that is obtained at leading order in the small $G_N$ expansion, and impose a UV cutoff $\Lam$. We may define a space, $\sH_\Om^\Lam,$  as the closure of the linear span of the states in $\sH_\Om^{\rm (Fock)}$ with energies below $\Lam$.\footnote{The precise way to impose the UV cutoff should not be important. Some other method can be used, especially in spacetime geometries where a bulk energy is not well-defined.}
Imposing such a UV cutoff will lead to nonlocality at the length scale $1/\Lam$. However, if we are only interested in physics at distance scales much larger than $1/\Lam$~(or energy scales much smaller than $\Lam$), this approximate description should be valid. We can then use $\sH_\Om^\Lam$ as an approximation of $\sH^{\rm (code)}_\Om$. 
Since states in $\sH_\Om^\Lam$ are those in $\sH_\Om^{\rm (Fock)}$, which are defined in the infinite $N$ limit, they are clearly different from those in $\sH^{\rm (code)}_\Om$, which are exact finite $N$ states.\footnote{There does not appear to be any canonical mapping between states in $\sH_\Om^\Lam$ and those in $\sH^{\rm (code)}_\Om$.} 
Nevertheless, our assumption is that for a sufficiently large $N$ a state in $\sH^{\rm (code)}_\Om$ can be well approximated by that in  $\sH_\Om^\Lam$, although at the moment we do not know how to precisely control the approximation, i.e., to quantify the errors. 

$\sH^{\rm (code)}_\Om \approx \sH_\Om^\Lam$ provides the finite $N$ (finite $G_N$) counterpart of the GNS Hilbert space (which is only defined in the $N \to \infty$ limit where it satisfies $\sH_\Om^{\rm (GNS)} = \sH_\Om^{\rm (Fock)}$). We should have  $\Lam \to \infty$ as $G_N \to 0$, so that $\sH_\Om^\Lam$ goes over to $\sH_\Om^{\rm (Fock)}$ in the $G_N \to 0$ limit.

After giving an approximate description of the code subspace $\sH^{\rm (code)}_\Om$, we need to specify the operators acting on the code subspace, i.e.,  the ``low-energy'' operators.  Since $\sH^{\rm (code)}_\Om$ involves exact quantum states at finite $N$, a precise description of operators  acting on $\sH^{\rm (code)}_\Om$ is also out of our control. Similar to our approximate description of $\sH^{\rm (code)}_\Om$, we will consider an approximate description motivated by the $N \to \infty$ limit. 
 In the $N \to \infty$ limit, the operators that act on $\sH_\Om^{\rm (GNS)}$ belong to an algebra, $\sY,$ that is generated by products of single-trace operators. 
 It is thus natural to use products of single-trace operators to approximate operators in $\sB (\sH^{\rm (code)}_\Om)$. Since the products of single-trace operators generate the full boundary algebra and since products of too many single-trace operators should not have an approximate semi-classical gravity description at finite $N$, we put a cutoff on the number of factors in a product. More explicitly, consider the vector space $\sY^K$ spanned by products of up to $K$ single-trace operators,
with $1\ll K \ll N$.
$K$ can be chosen such that $K \to \infty$ as $N \to \infty$ such that $\sY^K$ becomes the algebra of single-trace operators at infinite $N$.  
 
The finite $N$ counterpart of a bulk local operator $\phi (X)$ can be defined from~\eqref{eun0} as 
 \be \label{eq:bulkQGOp}
\hat \phi (X) =  \sO_{G} (X) = \int d^{d} x \, K \le(X; x\ri)  \sO (x),
\ee
where $\hat \phi (X)$ is now interpreted as an operator in the bulk quantum gravity theory at finite $G_N$. 

The operator ${\hat \phi}(X)$ defined in~\eqref{eq:bulkQGOp} will only approximately preserve $\sH^{\rm (code)}_{\Om}$. Its action on $\sH^{\rm (code)}_{\Om}$ can be further approximated by its action on $\sH_\Om^\Lam$ through the use of the operator $\pi_{\Om}(\sO(x))$ 
in lieu of $\sO(x)$ itself. Concretely, the action of ${\hat \phi}(X)$ on $\sH_{\Om}^{\rm (code)}$ at a large but finite $N$ should be approximated by the action of $\pi_{\Om}({\hat \phi}(X))$ on $\sH_{\Om}^{\Lam}$ though this approximation is uncontrolled and hard to explicitly describe due to the lack of a canonical map from $\sH_{\Om}^\Lam$ to $\sH_{\Om}^{\rm (code)}.$

\subsubsection{Entanglement wedge reconstruction} 

Now consider a spatial boundary subregion $R$. We have the local operator algebra $\sB^{(N)}_R$ which is the algebra of all operators localized in $R$. At a large but finite $N$, it is natural to introduce $\sY^K_{\hat R}$ as the set of operators in $\sY^K$ that are localized in $\hat R$ as a finite $N$ counterpart to $\sY_{\hat R}$.

We also need a finite $N$ counterpart of $\sX_R$, which can be defined as the subset of operators in $\sB^{(N)}_R$ that acts on the code space.
We can only approximately describe this algebra.
We say an operator $A \in \sX_R^\ep$ if $A \in \sB^{(N)}_R$ and there exists an operator $\tilde A \in \sY^K$ such that 
\be \label{eq:defApproxX}
\sup_{\ket{\psi} \in \sH_\Om^{\rm (code)}}  ||A \ket{\psi} - \tilde A \ket{\psi}||  < \ep  \ .
\ee
We should choose $\ep$ such that it goes to zero as $N \to \infty$. 
We emphasize that the element of $\sY^K$ whose action on $\sH^{\rm (code)}_{\Om}$ approximately equals that of $A \in \sB^{(N)}_R$, need not be localized in $\hat R$.
From the definition we have $\sY_{\hat R}^K \subseteq \sX_R^\ep$.  Neither $\sY_{\hat R}^K$ nor $\sX_R^\ep$ is an algebra.
Operators in $\sB^{(N)}_R$ that are not in $\sX_R^\ep$ do not act on $ \sH^{\rm (code)}_\Om$ in the sense they take a state in $ \sH^{\rm (code)}_\Om$ ``far away'' from it.

The above definition of $\sX_R^\ep$ is abstract as we do not have control over $ \sH^{\rm (code)}_\Om$. An approximate description can be given, again with hints from the story in $1/N$ perturbation theory, where $\sX_R$ can be considered as being generated by the large $N$ limit of modular-flowed operators: 
\be 
\sO (s, \vx) =  \De_{R}^{-is} \sO (0,\vx) \De_{R}^{is}, \quad  \vx \in R ,
\ee
where $\De_R$ is the modular operator associated with $\sB^{(N)}_R$ in the vacuum state $\ket{\Om}$. 
It is then natural to expect that operators in $\sX_R^\ep$ may be obtained from the modular flow of operators in $\sY^K_R$~(with $\ep$ and $K$ related in some way).

In $1/N$ perturbation theory, a bulk operator $\phi (X)$ in the entanglement wedge $\hat \fb_R$ of
$R$ can be reconstructed in the form~\cite{Jafferis:2015del,Faulkner:2017vdd} 
\bega \label{eq:exactModRecon}
	\phi(X) = \int_R d^{d-1}\vy \int_{-\infty}^{\infty} ds~ K_R\le(X; \vy, s\ri)  \tilde \De_{R}^{-is}\pi_{\Om}\le(\sO (0,\vy)\ri)\tilde \De_{R}^{is} = \pi_\Om (\sO_R (X)), \\
	\sO_R (X) \equiv  \int_R d^{d-1}\vy \int_{-\infty}^{\infty} ds~ K_R\le(X; \vy, s\ri)  \De_{R}^{-is} \sO (0,\vy) \De_{R}^{is}, 
	\label{eun1} 
\end{gather} 
where $K_R(X; \vy, s)$ is the modular reconstruction kernel for $R$ and $\tilde \De_{R}$ is the modular operator for $\sX_R$ in the state $\ket{\bid}_{\Om}$ (the GNS representation of $\ket{\Om}$ in $\sH^{\rm (GNS)}_{\Om}$). 
The second equality of~\eqref{eq:exactModRecon}, which follows from~\cite{Jafferis:2015del}, is a highly non-trivial statement 
regarding the large $N$ limit of modular flows associated with $\sB^{(N)}_R$. 
Exactly parallel equations to~\eqref{eq:exactModRecon}--\eqref{eun1} can be written down for any subregion $\tilde R$ satisfying $X \in \hat \fb_{\tilde R}$.

Note that both~\eqref{eun0} and~\eqref{eq:exactModRecon} are exact identities in the $1/N$ expansion, and 
thus all the infinite number of ways of reconstructing $\phi (X)$ in fact yield {\it the same} operator on $\sH^{\rm (GNS)}_{\Om}$. 
Despite the equalities 
\be \label{Ineq}
\pi_\Om (\sO_G (X)) = \pi_\Om (\sO_R (X)) =  \pi_\Om (\sO_{\tilde R} (X)) ,
\ee
$\sO_G (X)$, $\sO_R (X)$, and $\sO_{\tilde R} (X)$ are completely different operators at finite $N$. 
It is just that in the large $N$ limit, their representations on the GNS Hilbert space $\sH^{\rm (GNS)}_\Om$ coincide.

At a finite $N$ (non-perturbative in $G_N$), if $X \in \hat \fb_R$, then, recalling that $\hat \phi (X) = \sO_G(X),$ entanglement wedge reconstruction implies that there is an operator $A \in \sX^\ep_R$ such that
\be 
||\sO_G (X) \ket{\psi} - A \ket{\psi}|| < \ep, \quad \forall \ket{\psi} \in \sH^{\rm (code)}_\Om \ .
\ee
Modular reconstruction further suggests that we have $A = \sO_R(X)$ as in~\eqref{eun1} and therefore
\be \label{enn0}
||\sO_G (X) \ket{\psi} - \sO_R  (X) \ket{\psi}|| < \ep, \quad \forall \ket{\psi} \in \sH^{\rm (code)}_\Om \ .
\ee
The above equation can be considered as the finite $N$ version of~\eqref{Ineq}.

\subsubsection{Superadditivity}

With our ``approximate'' definition of $\sX_R^\ep$, the superadditivity property persists at a finite, but large, $N$ through 
\be 
\sX_{R_1}^\ep \tilde  \lor \sX_{R_2}^\ep \subsetneq \sX^\ep_{R_1 \cup R_2}, \quad  \sX^\ep_{R_1 \cap R_2} \subsetneq \sX^\ep_{R_1} \cap \sX^\ep_{R_2} ,
\ee
where the first inclusion can be understood as the existence of more low energy operations 
accessible in $R_1 \cup R_2$ than can be obtained through the combination of those of in $R_1$ and $R_2$. Since $\sX_{R}^\ep$ is no longer a closed algebra, the symbol $\sX^\ep_{R_1}\tilde  \lor \sX^\ep_{R_2}$ denotes the vector space of operators
obtained by taking polynomials of operators in $\sX^\ep_{R_1} \cup \sX^\ep_{R_2}$, and then truncating to those that approximately act on $\sH_\Om^{\rm (code)}$ in the sense of~\eqref{eq:defApproxX}.

In the bulk, we accordingly have 
\be 
\fb_{R_1} \cup \fb_{R_2} \subsetneq \fb_{R_1 \cup R_2}, \quad  \fb_{R_1 \cap R_2} \subsetneq \fb_{R_1} \cap \fb_{R_2}  \ .
\ee
However, quantum spacetime fluctuations at finite $N$ now make the definitions of entanglement wedges approximate, and thus the above relations are again only defined in an approximate sense. 

\subsection{Models from quantum error correction} \label{sec:QEC}

Motivated by local reconstruction and the superadditivity structure~\eqref{Aoo2} of entanglement wedges
many models of holography based on quantum error correction (QEC) have been proposed~\cite{Pastawski:2015qua, Hayden:2016cfa}.
In this section we focus on QEC models for ``small'' code subspaces, i.e. those whose dimension does not grow exponentially in $1/G_N$.

In a quantum error correcting code, a logical Hilbert space (code subspace) $\mathcal{H}_{\text{code}}$ is embedded into a physical Hilbert space $\mathcal{H}_{\text{phys}}$ through an isometry $V$, i.e.
 \be 
 V: \sH_{\rm code} \to \sH_{\rm phys} \ .
 \ee
 States in $\mathcal{H}_{\text{code}}$ are highly entangled and specifically ``designed" such that they and operations on them are protected from certain types of errors.
In the holographic context, 
one identifies the low-energy Hilbert space $\sH_{\rm bulk}$ around a bulk geometry as the code space $ \sH_{\rm code} $ and 
the boundary Hilbert space, $\sH_{\rm CFT}$, as $\mathcal{H}_{\text{phys}}$. QEC models of holography are therefore based on the existence of an isometry
\be\label{oeu}
 V: \sH_{\rm bulk} \to \sH_{\rm CFT} \ .
\ee

From the discussion around Fig.~\ref{fig:strH} a model for holography in terms of~\eqref{oeu} is meaningful 
only for a large finite $N$, with $N$ (or $G_N$) treated non-perturbatively. In $1/N$ (or $G_N$) perturbation theory 
 there is {\it equivalence} between the bulk Fock space and the boundary GNS Hilbert space, rather than embedding. 
 As discussed in section~\ref{sec:approxCodeSpace}, when $N$ is treated non-perturbatively, the formulation of $\sH_{\rm bulk}$~(or $\sH_{\rm code}$) can only be described approximately. Entanglement wedge reconstruction is also only approximate in this regime. In particular, closed algebras such as $\sY$, $\sX_R$ and $\sY_{\hat R}$ can no longer be defined. Instead, one has to deal with vector spaces of operators whose elements either cannot be explicitly described or only provide an (uncontrolled) approximate description of low-energy bulk physics.

Models based on exact quantum error correction provide precise definitions of objects analogous to those described in section~\ref{sec:EWRFinN}. Specifically, within any QEC model, closed algebras that are analogous to the $\sX^{\ep}_R$ of section~\ref{sec:EWRFinN} are explicitly described.
Such models do not exactly define the same structure of finite $N$ holography, but they nonetheless can offer conceptual insights into finite $N$ holography. 

Entanglement wedge reconstruction on a bulk time-slice 
has the following features:
\ben
\item Complementary recovery---the bulk subregions $\fb_R$ and $\bar \fb_R$ are respectively reconstrutable from boundary subregions $R$ and $\bar R$.
\item  Superadditivity of entanglement wedges/redundant reconstructions---a boundary region $R_1 \cup R_2$ can reconstruct a bulk subregion $\fb_{R_1 \cup R_2}$ that is larger than $\fb_{R_1} \cup \fb_{R_2},$ the union of regions reconstructed by $R_1$ and $R_2$. Bulk operators can be reconstructed on multiple boundary subregions.

\item Reeh-Schlieder---the bulk Hilbert space contains a state that is cyclic and separating for the algebra associated to any proper bulk subregion. The image of this state under the encoding isometry is a cyclic and separating state in the CFT Hilbert space for the full CFT algebra associated to any proper boundary subregion. 
\een

So far, no isometric QEC model appears to exhibit all features 1-3. In fact, there appears to be a fundamental obstruction to having 1-3 
simultaneously with exact quantum error correction~\cite{Kelly:2016edc, Faulkner:2020hzi}.\footnote{There are also arguments for the necessity of approximate QEC based on a ``large'' code spaces which include black hole states~\cite{Hayden:2018khn}, i.e., code spaces with dimension $e^{O(N^2)}$, which we do not consider here. 
}

Tensor network models~\cite{Pastawski:2015qua, Hayden:2016cfa}
provide instructive toy models which capture many aspects of holographic systems, including items 1 and 2 for certain subregions.\footnote{See~\cite{Swingle:2009bg} for an earlier model of holography using tensor networks motivated by the renormalization group.}  However, in tensor network models boundary and bulk subregions are described by finite dimensional type I algebras and are thus incompatible with the Reeh-Schlieder property for generic subregions.

Similar statements apply to the more abstract approach~\cite{Cotler:2017erl} based on universal recovery channels. 
Again the algebras associated to subregions are still taken to be finite-dimensional and the conclusion that approximate equality of relative entropy implies approximate entanglement wedge reconstruction is shown only in this finite-dimensional setting. Thus the abstract models considered in~\cite{Cotler:2017erl} does not satisfy the Reeh-Schlieder property for generic subregions.

\begin{figure}[!h]
\begin{center}
\includegraphics[width=6.5cm]{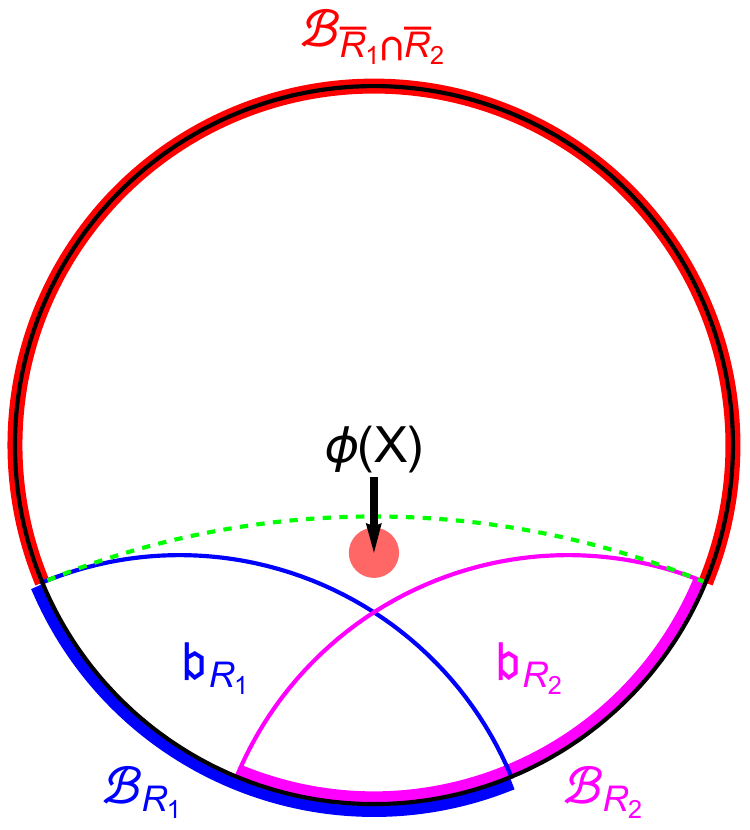}
\caption{ A configuration of overlapping intervals that shows that complementary recovery and the Reeh-Schlieder theorem are inconsistent with the holographic interpretation of algebraic QEC codes. 
Consider a bulk operator outside both $\fb_{R_1}$ and $\fb_{R_2}$, such as $\phi(X)$, which can be reconstructed from both $\sB_{\bar R_1}$ and $\sB_{\bar R_2}$.
It can be shown that for an exact QEC, the existence of a separating state for $\sB_{\bar R_1 \cup \bar R_2}$ implies that $\phi (X)$ is an element of $\sB_{\bar R_1 \cap \bar R_2}$. Thus it cannot be in the entanglement wedge of $R_1 \cup R_2$, showing that the entanglement wedge in such models is additive. This is inconsistent with holography since the RT formula suggests that $R_1 \cup R_2$ can reconstruct all operators up to the green dashed curve in the figure, rather than only those within the blue or magenta curves.
}
\label{fig:reehSchlieder}
\end{center}
\end{figure}

An alternative approach is to assume that a QEC code exists and consider its properties at an abstract level. 
In this approach, one can consider general von Neumann algebras 
where there can exist cyclic and separating states for all proper subalgebras of the system.
However it can be shown that complementary recovery and the existence of 
a separating state for the algebra of a particular boundary subregion imply that redundant descriptions of bulk operators (or superadditivity of entanglement wedges) cannot happen~\cite{Kelly:2016edc, Faulkner:2020hzi}, failing to capture feature 2.
This implies that entanglement wedges in such abstract QEC models are additive: 
\be 
\fb_{R_1} \cup \fb_{R_2} = \fb_{R_1 \cup R_2} \ .
\ee
See figure~\ref{fig:reehSchlieder}.

Therefore it appears that, even for a small code subspace, to reproduce all three features of entanglement wedge reconstruction, quantum error correction has to be approximate; however, it is not yet clear what type approximate QEC should be used. 

Another alternative is to give up on the existence of an isometric embedding of the bulk into the boundary Hilbert space. A recent discussion along these lines appeared in~\cite{Faulkner:2022ada}. In that work, Faulkner and Li consider embedding the GNS Hilbert space, defined in the $N \to \infty$ limit and equal to the bulk Fock space, into the Hilbert space of the CFT at finite $N$. These embeddings are not isometric, but it is argued that the embeddings approach an isometry as $N \to \infty$.
Such an asymptotically isometric code can satisfy all three features discussed above and appears to be a promising way to capture holography at a large but finite value of $N$ using QEC.

Finally, we note that the above discussion has only considered a single bulk time-slice and its asymptotic limit to the boundary. There are additional difficulties using models based on QEC when ones tries to capture time evolution and causal structure.\footnote{See~\cite{Faist:2019ahr} for discussion on the difficulties adding non-trivial time-evolution in tensor network models.} 
The main issue is that genuine boundary time-evolution does not preserve the code subspace. Namely, the genuine boundary Hamiltonian has a form $H_{\rm bdry} = \sum_{\alpha} H_{\alpha}$ where $H_{\alpha}$ are local boundary operators that may include few-body interactions. For such a Hamiltonian, we generically have
$e^{iH_{\rm bdry}t}\ket{\psi} \notin \sH_{\rm code}$ for $\ket{\psi} \in \sH_{\rm code}$. 
In contrast, the bulk Hamiltonian, by definition, preserves the code subspace. Suppose that we have a local bulk theory so that the bulk effective field theory Hamiltonian can be written as $H_{\rm bulk} = \sum_i h_i$ for local bulk terms $h_i$. 
Under the embedding isometry, this maps to a boundary operator $VH_{\rm bulk} V^{\dagger} = \sum_i V h_i V^{\dagger}.$ This operator preserves the code subspace of the boundary theory but is highly non-local and is not related to the genuine boundary Hamiltonian in any simple way.

\section{Implications of superadditivity (II): Performance of quantum tasks} \label{sec:cwt}

Superadditivity for large $N$ algebras provides an interpretation of the connected wedge theorems (CWTs) recently discussed in~\cite{May:2019yxi, May:2019odp, May:2021nrl, May:2022clu}. The CWTs concern the duality between quantum tasks that can be completed by bulk and boundary protocols. 
May~\cite{May:2019yxi} found that certain quantum tasks cannot be performed through a local boundary protocol but can be performed through a local bulk protocol.\footnote{Recently the CWT has been extended to finite cutoff AdS~\cite{Mori:2023swn} and the static patch of de Sitter~\cite{Franken:2024wmh} using a notion of ``induced causlity.'' In this paper, we assume standard causality for the boundary manifold so our results are not directly applicable to the settings of~\cite{Mori:2023swn} and~\cite{Franken:2024wmh}.} The existence of the bulk protocol and holographic duality of quantum tasks implies the existence of a non-local boundary protocol to perform the task. The existence of a non-local boundary protocol was then interpreted as being facilitated by a large amount of mutual information between certain configurations of disjoint boundary intervals.

{Here we provide an alternative interpretation of the non-local boundary protocol as being due to the superadditivity of certain boundary algebras. This re-casts the existence of the protocol in terms of the existence of non-trivial operators: in the large $N$ limit, there are more nonlocal operations available than one would have naively expected based on additivity. The reinterpretation also leads to a natural reformulation and generalization of the CWTs. Finally, since the bulk construction of the protocol is based on the causal structure of the bulk geometry, the CWTs can be viewed as establishing new (and rather subtle) connections between the superadditivity of  boundary CFT algebras and the causal structure of the bulk spacetime. }

Below we will first review the two-to-two theorem of~\cite{May:2019yxi, May:2019odp, May:2021nrl, May:2022clu} in AdS$_3$,\footnote{We believe the discussion can be extended to the $n$-to-$n$ theorem of~\cite{May:2022clu} and leave this to future work.} then we present our generalization in terms of superadditivity, and finally we elaborate on the generalization. 

\subsection{The two-to-two theorem}\label{sec:22}

Consider a bulk spacetime with a single asymptotically AdS$_3$ region which is dual to a boundary CFT$_2$ on $\mathbb{R} \times S^1$. We denote boundary points with lowercase latin letters with subscripts, e.g. $c_1,c_2,r_1,r_2,$ etc. The set of all boundary points that can be reached by a future-directed (past-directed) causal curve {\it on the boundary} from a boundary set $R$ is denoted by $\hat{J}^+(R)$ ($\hat{J}^-(R)$). Similarly, the set of all bulk points that can be reached by a future-directed (past-directed) causal curve {\it in the bulk} from a bulk set $\fr$ is denoted by $J^+(\fr)$ ($J^-(\fr)$).

\begin{figure}[h]
\begin{center}
\includegraphics[width=6.5cm]{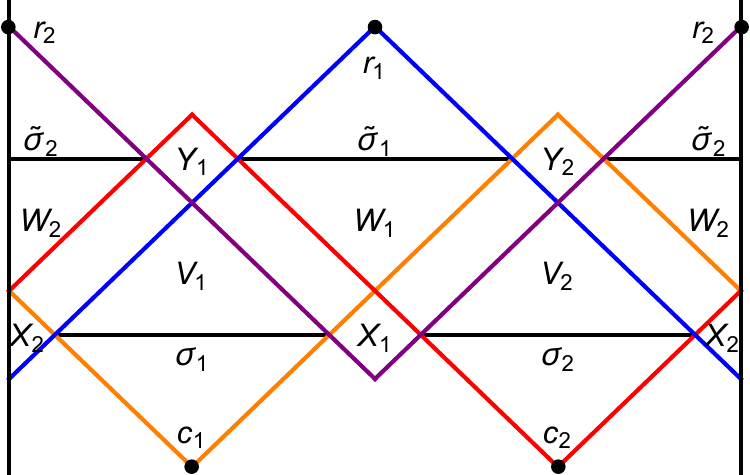}
\caption{The set-up for the two-to-two theorem on $\mathbb{R} \times S^1$ with time running vertically and the solid black vertical lines identified. We consider two input points ($c_1, c_2$) and two output points ($r_1, r_2$) such that information from both input points can be locally processed and then sent to {\it either} output point, but not to both output points simultaneously.}
\label{fig:connWedge}
\end{center}
\end{figure}

By a bulk-only scattering configuration of boundary input points $c_1,c_2$ and output points $r_1,r_2$, we mean that the boundary scattering region $\hat{J}^+(c_1)\cap \hat{J}^+(c_2)\cap \hat{J}^-(r_1)\cap \hat{J}^-(r_2)$ is empty, but the corresponding bulk scattering region 
\be\label{des0}
S_0 \equiv {J}^+(c_1)\cap {J}^+(c_2)\cap {J}^-(r_1)\cap {J}^-(r_2)
\ee
is non-empty. 
It is convenient to introduce boundary 
{\it spacetime} subregions defined by $V_i = \hat{J}^+(c_i) \cap \hat{J}^-(r_1) \cap \hat{J}^-(r_2)$ and $W_i = \hat{J}^+(c_1) \cap \hat{J}^+(c_2) \cap \hat{J}^-(r_i),$ respectively. See figure~\ref{fig:connWedge}. $V_1$ can be interpreted as the spacetime region where local operations can be used to perform tasks with input from $c_1$ and outputs to $r_1, r_2$, while $W_1$ is the spacetime region where local operations can be used to perform tasks with inputs from $c_1$ and $c_2$, with output to $r_1$. $V_2$ and $W_2$ have analogous interpretations. The non-existence of a boundary scattering region implies that there is no intersection among any of these regions. Furthermore, all of these boundary subregions are diamonds, i.e. boundary domains of dependence of some interval~\cite{May:2022clu}. For example,  $V_1$, $V_2,$ $W_1,$ $W_2$ are respectively the domains of dependence of the
intervals $\sigma_1$, $\sigma_2,$ $\tilde{\sigma}_1,$ $\tilde{\sigma}_2$.
An important feature in two dimensions is that the (boundary) causal complement of $V_1 \cup V_2$ is also the union of two diamonds. We denote this by $X_1 \cup X_2 = (V_1 \cup V_2)',$ where each of $X_{1,2}$ is a single diamond. Similarly we can define $Y_1 \cup Y_2 = (W_1 \cup W_2)',$ where $Y_{1,2}$ are single diamonds.

It is convenient to consider the following {generalization} of $S_0$, 
\be \label{eq:defOfGenScattReg}
	S \equiv J^+(\hat \fb_{V_1})\cap J^+(\hat \fb_{V_2})\cap J^-(\hat \fb_{W_1})\cap J^-(\hat\fb_{W_2}) 
\ee
which is non-empty whenever $S_0$ is non-empty, i.e. $S_0 \subseteq S$. 
The CWT of~\cite{May:2022clu} is the following:  suppose that the bulk satisfies the null curvature condition and that the RT surface can be computed by the maximin prescription, then non-empty $S$ implies that {the entanglement wedge of $V_1 \cup V_2$ must be connected}.
The same statement applies to the entanglement wedge of $W_1 \cup W_2$. 

Some remarks on the theorem:

\ben

\item Connectedness of the entanglement wedge of $V_1 \cup V_2$ can be interpreted on the boundary as the system having a large amount  (i.e. $O(1/G_N)$) of mutual information between $V_1$ and $V_2$~\cite{May:2019yxi}. For certain cases there exist explicit boundary protocols that make use of the large amount of mutual information  as a resource to perform tasks~\cite{May:2019yxi}; however, it is unclear if there is a sense in which those non-local boundary protocols are dual to the local bulk protocol.

From our discussion of Sec.~\ref{sec:addAnom}, we see that the connectedness of the entanglement wedge can also be interpreted in terms of superadditivity for the involved algebras, i.e. 
\be 
\sX_{V_1} \lor \sX_{V_2} \subsetneq \sX_{V_1 \cup V_2}  ,
\ee
which in turn implies that there are additional nonlocal operations in $V_1 \cup V_2$ that cannot be obtained by combining those of $V_1$ and $V_2$.

\item Since $S$ does not have a direct boundary interpretation, the theorem cannot be formulated as an intrinsic boundary statement.

\item The converse statement does not hold~\cite{May:2019odp}. In other words, the theorem is not an equivalence statement: while the existence of a bulk scattering region requires the system to have superadditivity, having superadditivity is not enough to ensure that quantum tasks can be performed.

This last point is somewhat puzzling. Recall that $V_1$ ($V_2$) is the boundary spacetime region that can directly connect $c_1$ ($c_2$) to $r_1, r_2$. 
Connectedness of the entanglement wedge of $V_1 \cup V_2$ suggests that 
$V_1 \cup V_2$ should be most naturally thought of as a single object rather than being formed out of disjoint pieces. This is clear in the bulk since there is no canonical way to separate $\fb_{V_1 \cup V_2}$ into two pieces that are individually associated with $V_1$ and $V_2$. We can view this fact as a bulk geometrization of a `quantum information connection' between $V_1$ and $V_2$.
\footnote{This can potentially be made precise by noting that $\sX_{V_1 \cup V_2}$ is generated by the large $N$ limit of operators of the form $\De_{V_1\cup V_2, \Omega}^{is} \sO(x) \De_{V_1\cup V_2, \Omega}^{-is}$ for $x \in \sigma_1 \cup \sigma_2$ and $s \in \mathbb{R}$. For any $s\neq 0$ the resulting operator is a highly non-trivial mixture of operators in $V_1 \cup V_2$. This operator may be separated into pieces individually associated with $V_1$ and $V_2$ at any finite $N$; however, these individual pieces do not survive the large $N$ limit on their own and thus, at $N = \infty$, one can only view the operator as associated to $V_1 \cup V_2$.} 
The `extra' operations in $V_1 \cup V_2$ should imply that there is a way to perform quantum tasks efficiently through the bulk, and thus appears to be in tension with the lack of a converse for the theorem.\footnote{See~\cite{Caminiti:2024ctd} for a discussion of the CWT in BTZ and AdS$_3$ defect geometries where a different modification of the CWT allows the converse to hold.}

\een
Below we will present a reformulation/generalization of the theorem that will have an intrinsic boundary formulation. Furthermore, it is conjectured to be an equivalence statement, and has a clear physical interpretation, resolving the tension mentioned above.

\subsection{Generalizing the CWT to an equivalence statement} \label{sec:genCWT}

We define a new generalized ``scattering'' region as
\be \label{eq:EintE}
	S_E \equiv \hat \fb_{V_1 \cup V_2} \cap \hat \fb_{W_1 \cup W_2}  \ .
\ee
In general $S_E \neq S,$ but assuming the null curvature condition we always have (see Appendix~\ref{app:SEcontS} and~\cite{May:2019odp})\footnote{This can be deduced by noting that the argument that $\hat \fb_{V_1 \cup V_2}$ always contains the scattering region which is given in~\cite{May:2019odp} can be equally well applied to $\hat \fb_{W_1 \cup W_2}$. See appendix~\ref{app:SEcontS}.}
\be 
	S \subseteq S_E  \ .
\ee
Thus $S_E$ can be non-empty when $S$ is empty, but not vice-versa. 
$S_E$ being non-empty implies quantum information contained in $V_1 \cup V_2$ can be directly transferred through the bulk 
to $W_1 \cup W_2$. It should thus also lead to efficient operations to perform quantum tasks from $c_1, c_2$ to $r_1, r_2$. 

Our proposal for the generalized CWT is then the equivalence between the existence of `extra' operations in $V_1 \cup V_2$ and the existence of non-local protocols for quantum tasks from $c_1,~c_2$ to $r_1,~r_2$.
Mathematically, this is stated as follows. 
\medskip

\noindent {\bf Conjecture for a Generalized Connected Wedge Theorem (GCWT)}: \\
Let $V_1, V_2, W_1, W_2$ be the boundary spacetime subregions (on $\mathbb{R} \times S^1$) obtained from a configuration of boundary input points, $c_1, c_2,$ and output points, $r_1, r_2,$ such that all of $V_1, V_2, W_1, W_2$ are non-empty, but no boundary scattering region exists. Assume also that the bulk satisfies the null curvature condition and that there is a bulk Cauchy surface $\Sigma_V,$ to the past of $\hat \fb_{V_1 \cup V_2},$ and another bulk Cauchy surface $\Sigma_W,$ to future of $\hat \fb_{W_1 \cup W_2},$ such that the region between $\Sigma_V$ and $\Sigma_W$ has no singularities.\footnote{It is possible that the assumption of no singularities is unnecessary and can be weakened.} Then $S_E$ is non-empty {\it if and only if} $\fb_{V_1 \cup V_2}$ is connected.

In the forward direction, this is a strengthening of the CWT. It says that we only need $S_E$ to be non-empty (which is a weaker condition than non-empty $S$) for a connected $\fb_{V_1 \cup V_2}$. In the backward direction, this is a generalization of the converse of the CWT. Importantly, $S_E$ can be bigger than $S$, so, as we will see below, the counter-examples to the converse of the CWT given in~\cite{May:2019odp} are not counter-examples to our proposed generalized CWT. Those examples have a connected entanglement wedge and empty $S$, but do not have empty $S_E.$

We can re-phrase the proposed theorem in the language of boundary algebras as follows. 
\medskip

\noindent {\bf Conjecture for Superadditivity in Holographic CFT$_2$ as $\mathbf{N \to \infty}$}\\
Suppose that we have a configuration of (non-empty) boundary diamonds $V_1,V_2,W_1,W_2$ defined by a boundary two-to-two scattering problem on $\mathbb{R} \times S^1$ with no boundary scattering region. 
Then, working about any pure boundary state such that the net of large $N$ algebras $\{\sX_R\}$ obeys causality, we have the equivalence
\be \label{eq:genCWT}
	\sX_{V_1 \cup V_2} \cap \sX_{W_1 \cup W_2} \neq \emptyset \quad \Longleftrightarrow \quad \sX_{V_1 \cup V_2} \supsetneq \sX_{V_1} \vee \sX_{V_2} \ ,
\ee
where the inclusion on the RHS is proper.\\

The statement~\eqref{eq:genCWT} is a pure boundary statement and thus can be taken as a statement about the large-$N$ limit (about a pure state) of holographic CFTs. In the forward direction, this theorem says that if there are operators that are elements of both the algebra associated to $V_1 \cup V_2$ and the algebra associated to $W_1 \cup W_2$ (which are disjoint but causally related boundary subregions) then the algebra of $V_1 \cup V_2$ cannot be additive, i.e. there are `extra' operators in the algebra of $V_1 \cup V_2$ that cannot be obtained additively from the algebras of $V_1$ and $V_2$ separately. In the reverse direction, this says that when the algebra of $V_1 \cup V_2$ exhibits superadditivity, this algebra will have a non-trivial intersection with the algebra of $W_1 \cup W_2,$ i.e. the presence of `extra' operators in the algebra of $V_1 \cup V_2$ guarantees that there will be some operators that are common to $V_1 \cup V_2$ and $W_1 \cup W_2.$ \\

This statement makes sense physically. As discussed in item 3 at the end of Sec.~\ref{sec:22}, when the entanglement wedge of $V_1 \cup V_2$ is connected it should already mean that it is possible to perform nonlocal protocols for quantum tasks. Connectedness of the entanglement wedge for $V_1 \cup V_2$ means that we can perform non-local operations, while non-emptiness of $S_E$ is the algebraic statement that quantum tasks involving input at $c_1$ and $c_2$ and output to $r_1$ and $r_2$ can be performed using the bulk. 
The generalized CWT conjectures that the two are indeed equivalent. 

In the special case of the vacuum state, the GCWT can be proven, see appendix~\ref{app:vacGCWT}. Studying this special case highlights how the existence of operators that are not generated additively from $V_1$ and $V_2$ should be equivalent to the existence of a generalized bulk scattering region.

\subsection{Argument for the GCWT for a general spacetime}

In this subsection we first show that the counterexample to the converse of the CWT discussed in~\cite{May:2019odp} is avoided for the GCWT, and then we offer some arguments for the GCWT for general asymptotically AdS$_3$ spacetimes.

A counter-example to the converse of the CWT was given in~\cite{May:2019odp}, so an important check of the GCWT or its boundary formulation~\eqref{eq:genCWT} is that it evades the counter-example, which we now recall. 

Suppose we begin with the following configuration of boundary points,
\begin{equation} \label{eq:MTspecialInpOutPts}
\begin{aligned} 
	&c_1: \le(t = -{w\ov 2}, \theta = \pi\ri),~ c_2: \le(t = -{w\ov 2}, \theta = 0\ri), \\
	 &r_1: \le(t = {\pi + w\ov 2}, \theta = -{\pi \ov 2}\ri),~ r_2: \le(t = {\pi + w\ov 2}, \theta = {\pi \ov 2}\ri) \ .
\end{aligned}
\end{equation}
The boundary subregions $V_1,V_2$ are then the domains of dependence of disjoint angular intervals of width $w$ on the $t=0$ slice. Similarly, $X_1,X_2$ are domains of dependence of disjoint angular intervals of width $\pi-w$ on the $t=0$ slice. At $w={\pi \ov 2},$ this configuration of intervals then clearly possesses symmetry under ${\pi \ov 2}$ rotations with such rotations exchanging $V_1 \cup V_2$ and $X_1 \cup X_2.$ About the AdS-vacuum the entanglement wedge of $V_1 \cup V_2$ is ambiguous as we are sitting right at the phase transition: if $w$ increases slightly the entanglement wedge is connected, whereas if $w$ decreases slightly the entanglement wedge is disconnected. 

Suppose we add some spherically symmetric {and time-translation invariant} positive-energy dust to the bulk in such a way that the dual boundary state is still pure. The bulk dust configuration should be chosen so that the candidate extremal surfaces do not change. The presence of such positive energy will delay light rays from $c_{1,2}$ causing the scattering region to disappear. However, under an infinitesimal increase in the parameter $w$ characterizing the boundary configuration (i.e. $w \to {\pi \ov 2}+ \ep$) the entanglement wedge of $V_1 \cup V_2$ will become connected. The volume of the scattering region $S$ will be a continuous function of $w$ and thus will remain empty. This is thus a counterexample to the converse of CWT~\cite{May:2019odp}: one has a situation in which $\fb_{V_1 \cup V_2}$ is connected but $S$ is empty.

Now consider the GCWT~\eqref{eq:genCWT} in the same set-up with positive energy dust and $w = {\pi \ov 2} + \ep$. We still have a connected entanglement wedge for $V_1 \cup V_2,$ so the RHS of~\eqref{eq:genCWT} is satisfied.  If~\eqref{eq:genCWT} is to hold, the intersection of entanglement wedges of $V_1 \cup V_2$ and $W_1 \cup W_2$ must be non-empty. Because the boundary state is pure, the entanglement wedges of $V_1 \cup V_2$ and $W_1 \cup W_2$ are the bulk causal complements of those of $X_1 \cup X_2$ and $Y_1 \cup Y_2,$ respectively. We will construct $\hat \fb_{V_1 \cup V_2}$ as the bulk causal complement of $\hat \fb_{X_1 \cup X_2}$.  

Let $\ga_{X_1},~\ga_{X_2}$ denote the two pieces of the RT surface bounding (the disconnected bulk subregion) $\fb_{X_1 \cup X_2}.$ The future/past boundary of $\hat \fb_{V_1 \cup V_2}$ is then the part of the null surface generated by outgoing (with respect to $\fb_{X_1 \cup X_2}$) future-/past-directed lightrays orthogonal to $\ga_{X_1}$ and $\ga_{X_2}$ up until the rays from $\ga_{X_1}$ meet those from $\ga_{X_2}$. Compared to propagation in the vacuum, the lightrays that generate these null surfaces will be delayed by the presence of the positive-energy dust so rays from $\ga_{X_1}$ and $\ga_{X_2}$ will not meet until a later time compared to the vacuum case. Since the extremal surfaces $\ga_{X_{1,2}}$ have not changed by the inclusion of dust, the delay of these null rays fired from them leads to $\hat \fb_{V_1 \cup V_2}$ being a {\bf larger} bulk subregion compared to the region bounded by these null surfaces in the vacuum. 
The same is true of $\hat \fb_{W_1 \cup W_2},$ since rays from $\ga_{Y_{1,2}}$ will also be delayed by the positive energy. 

In the set-up we are considering, for $w = {\pi \ov 2} + \ep$ the entanglement wedge is connected and we know that {\it about the vacuum} the intersection $\hat \fb_{V_1 \cup V_2} \cap \hat \fb_{W_1 \cup W_2}$ is non-empty. However, we have just argued that the presence of postive-energy dust causes these entanglement wedges to become larger and consequently, their intersection remains non-empty in the presence of positive-energy dust.
Thus, in the set-up we have just discussed the entanglement wedge of $V_1 \cup V_2$ is non-empty, but because $S_E$ is larger than $S$ we have found that $S_E$ remains non-empty, so this set-up satisfies the GCWT~\eqref{eq:genCWT}.

We will now explain more generally the differences between $S$ and $S_E$ that potentially allow \eqref{eq:genCWT} to hold, even though it does not hold when $S_E$ is replaced by $S$.

\begin{figure}[h]
        \centering
        \begin{subfigure}[b]{0.45\textwidth}
            \centering
		\includegraphics[width=\textwidth]{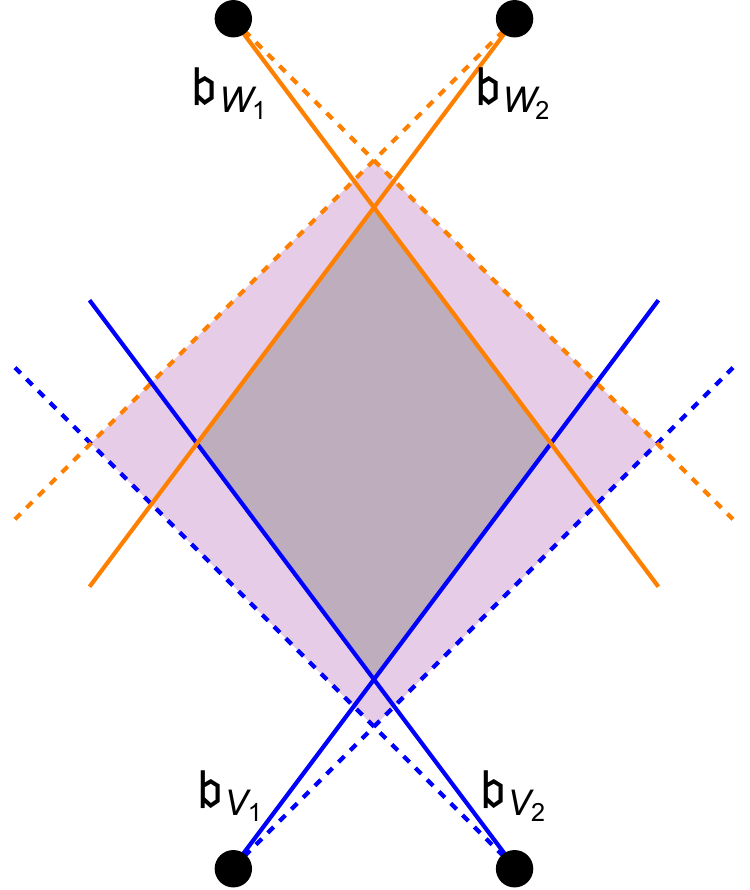} 
            \caption[]%
            {{\small} Shrinking of $S$ in the presence of positive-energy dust. }    
        \end{subfigure}
        \hfill
        \begin{subfigure}[b]{0.45\textwidth}   
            \centering 
		\includegraphics[width=\textwidth]{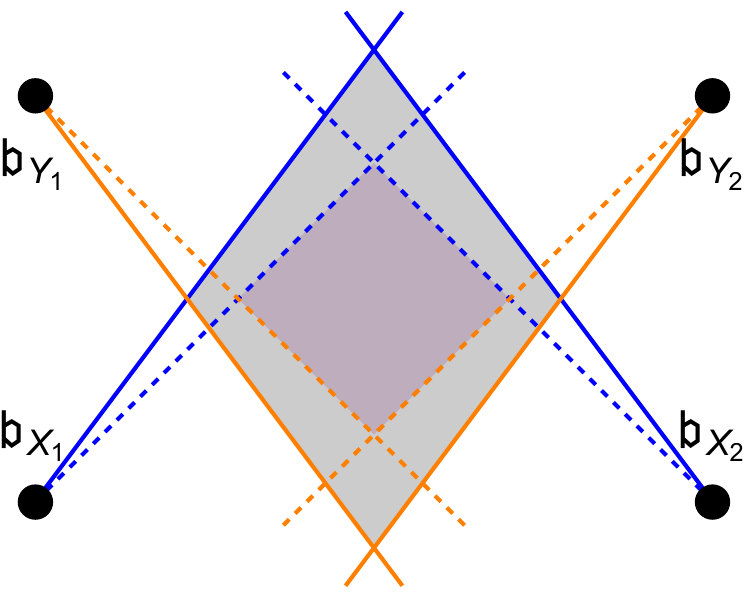}
            \caption[]%
            {{\small Growth of $S_E$ in the presence of positive-energy dust.}}    
        \end{subfigure}
        \caption[  ]
        {\small  Behavior of $S$ and $S_E$ in the presence of positive-energy dust. In each case, the propagation of light rays in vacuum is indicated by the dashed lines, while the propagation of light rays in the presence of dust is schematically shown by the solid lines.   } 
\label{fig:SvsSE}
\end{figure}

$S$ is the region to the future of the future-directed outgoing (with respect to $\fb_{V_{1,2}}$) null surfaces emanating from $\ga_{V_{1,2}}$ and to the past of the past-directed outgoing (with respect to $\fb_{W_{1,2}}$) null surfaces emanating from $\ga_{W_{1,2}}.$ The introduction of positive-energy matter delays all of these null rays causing this region between the null surfaces to shrink. See figure~\ref{fig:SvsSE} (a).
The loss of the scattering region $S$ in the counter-example of~\cite{May:2019odp} is then seen to be a special case of the fact that $S$ shrinks under the presence of positive energy in the bulk due to the associated delays of signals sent from or received in various bulk subregions. 

In a pure state, $\fb_{\bar R} = \overline{\fb_R}$. When $\fb_{V_1 \cup V_2}$ is connected, $\fb_{X_1 \cup X_2} = \fb_{X_1} \cup \fb_{X_2}$ is disconnected. $S_E$ can then be understood as the region to the {\it past} of the future-directed outgoing (with respect to $\fb_{X_{1,2}}$) null surfaces emanating from $\ga_{X_{1,2}}$ and to the {\it future} of the past-directed outgoing (with respect to $\fb_{Y_{1,2}}$) null surfaces emanating from $\ga_{Y_{1,2}}.$ 
Time-delays in the propagation of null rays therefore increase the size of the region. See figure~\ref{fig:SvsSE} (b).

We have argued that the inclusion $S_E \supseteq S$ can allow for a connected $\fb_{V_1 \cup V_2}$ to imply a non-empty $S_E.$ However, one may then worry that the antecedent of the CWT has been weakened too much such that the forward direction is no longer true. Specifically, one might worry that the region $S_E$ is too large and its non-emptiness can no longer imply a connected $\fb_{V_1 \cup V_2}$. A potential way to construct such a violation of~\eqref{eq:genCWT} is to take the counter-example set-up of~\cite{May:2019odp} and instead of increasing $w,$ slightly decrease it to $w = {\pi \ov 2} - \ep,$ so that the entanglement wedge of $V_1 \cup V_2$ is disconnected. 
The GCWT would then be violated if $S_E \neq \emptyset.$ However, we now have $\fb_{V_1 \cup V_2} = \fb_{V_1} \vee \fb_{V_2}$ (and $\fb_{W_1 \cup W_2} = \fb_{W_1} \vee \fb_{W_2}$), so that $\fb_{V_1 \cup V_2} \cap \fb_{W_1 \cup W_2}$ is empty and~\eqref{eq:genCWT} is not violated.

\section{Discussion} \label{sec:disc}

In this paper we have investigated the behavior of local algebras generated by quantum fields in the limit where the number of local degrees of freedom becomes infinite. We have argued that the structure of local algebras can undergo dramatic changes in this limit, most notably transitioning from a theory in which the additivity property holds to one in which it does not. We have demonstrated that this phenomenon occurs in both field theoretic examples in which the large $N$ limit does not have a local equation of motion as well as in holographic field theories. In the holographic context, we demonstrated that superadditivity can also have interesting consequences for black hole physics. Specifically, we showed that there are cases in which portions of the black hole interior may be reconstructed from the joint algebra of disjoint boundary subregions, even though no interior operators lie in the additive algebra.

We also argued that superadditivity of large $N$ algebras (i.e. the existence of configurations obeying~\eqref{hen3}) explains the success of quantum error correcting codes as toy models of holography. This is because such codes mimic the fact that the large $N$ algebras which describe semi-classical bulk physics contain operators which are not locally generated on the boundary and are responsible for the superadditivity of entanglement wedges~\eqref{Aoo2} in the bulk. 
We have also assessed the success of various QEC models.

Finally, we argued that superadditivity provides a new perspective on holographic quantum tasks. In particular, the existence of non-local boundary protocols is captured by the existence of non-locally generated operators in the large $N$ limit of holographic theories. This perspective allowed for a conjectured generalization of the connected wedge theorems for which the converse should hold and moreover, the generalization can be phrased as a pure boundary statement about the large $N$ limit of holographic CFTs. The proposed generalization resolves the tension between the existence of non-local operations and the ability to perform quantum tasks in the bulk due to the lack of the converse of the original CWTs as it instead proposes that the existence of non-locally generated operators and the ability to perform quantum tasks by a non-local protocol are equivalent.

There are many open questions remaining regarding superadditivity in large $N$ field theories, a few of which we mention below. Of particular interest is to understand the physical nature of the non-locally generated operators. Superadditivity in these field theories is more pronounced than in those studied in~\cite{Casini:2019kex, Casini:2021zgr} and the nature of the non-locally generated operators is not completely understood. It appears that the non-locally generated operators in this context are likely related to a difference in the modular structure of the corresponding algebras.

It is also of interest to understand how holography provides a measure for superadditivity, characterizing how many non-locally generated operators appear in each algebra.

Finally, we have demonstrated that superadditivity gives an explicit description of the operators that allow the boundary theory to perform quantum tasks non-locally. For possible applications to quantum information, we need to understand 
how to describe these operators at a large but finite $N$, which is clearly of great interest.

\vspace{0.2in}   \centerline{\bf{Acknowledgements}} \vspace{0.2in}
We would like to thank  Jacqueline Caminiti, Netta Engelhardt, Thomas Faulkner,  Daniel Harlow, Stefan Hollands, Jonah Kudler-Flam, Nima Lashkari, Juan Maldacena, Alex May, Hirosi Ooguri, Gautam Satishchandran, Jonathan Sorce, and Edward Witten for discussions. 
This work is supported by the Office of High Energy Physics of U.S. Department of Energy under grant Contract Number  DE-SC0012567 and DE-SC0020360 (MIT contract \# 578218). S.L. is supported by the Martin A.~and Helen Chooljian Member Fund and the Fund for Natural Sciences at the Institute for Advanced Study and the National Science Foundation under the grants PHY-2207584 and PHY-2209997.

\appendix

\section{Review and generalization of Araki's argument for additive algebras} \label{app:araki}

\subsection{Overlapping regions in the vacuum of CFT$_d$} \label{app:arakiVacArg}

We now review an argument due to Araki~\cite{ArakiTT} that shows that the {\it additive} algebra associated to a non-trivial union of double cones (for $d > 2$) is at least as large as that associated to its causal completion.\footnote{Our review here is informal and deals with a special case of those considered by Araki. We refer readers interested in the most general statement of Araki's result and the rigorous proof to the original paper~\cite{ArakiTT}.} The argument we describe applies equally well for $d=2;$ however, due to the properties of unions of double-cones in two dimensions, the argument only establishes equivalence of the additive algebra with the algebra of a region that is still smaller than the causal completion of the union of double-cones.

Araki's argument allows us to understand the nature of the (weak closure of the) algebra of operators generated by quantum fields restricted to the union of overlapping double-cones $\hat{R}_- \cup \hat{R}_+,$ where $\hat{R}_{\pm}$ are defined in~\eqref{eq:defDoubleConesPM}, in the vacuum sector of a relativistic QFT. We denote this algebra by $\sY^{\rm (vac)}_{\hat{R}_- \cup \hat{R}_+}.$

Consider a bounded operator, $C,$ that commutes with all locally generated operators in $\hat{R}_- \cup \hat{R}_+,$ i.e. $C \in \le(\sY^{\rm (vac)}_{\hat{R}_- \cup \hat{R}_+}\ri)'.$ Fixing two vectors $\ket{\Psi_{1,2}}$ in the dense domain of the Hilbert space upon which the local fields, $\sO(x),$ are defined as operator-valued distributions, we define the distribution 
\be \label{eq:commFcn}
	c(x) \equiv \bra{\Psi_1} C \sO(x) - \sO(x) C \ket{\Psi_2} \ .
\ee
From the definition of $C,$ the distribution $c(x)$ vanishes for $x \in \hat{R}_- \cup \hat{R}_+,$ and we now wish to show that it actually vanishes for any $x$ in a larger region $R_U^{\rm (vac)}.$\footnote{$R_U^{\rm (vac)}$ is a region whose causal domain is $\widehat{\fc_{R_-}\cup \fc_{R_+}}$ and is obtained by sending light light rays from $\fc_{R_-}\cup \fc_{R_+}$ to the boundary as described in section~\ref{sec:2din}.
} 
In other words, we show
\be 
\le(\sY^{\rm (vac)}_{\hat{R}_- \cup \hat{R}_+}\ri)'' = \sY^{\rm (vac)}_{R_U^{\rm (vac)}} \ .
\ee
We will see that $R_U^{\rm (vac)}$ coincides with $\hat{R}_+ \vee \hat{R}_- \equiv \le(\hat{R}_- \cup \hat{R}_+\ri)'' = \widehat{R_- \cup R_+}$ when $d>2$, while for $d=2$ it is the blue region given in Fig~\ref{fig:overint} (a).

The first step of Araki's argument  is to extend the distribution $c(x)$ on $d-$dimensional Minkowski space to a distribution on $(d+1)-$dimensional Minkowski space that obeys the wave equation. The extended distribution is defined by
\be 
\label{eq:extendFcn}
	c_{ext}(x,s) = \int \frac{d^d p}{(2\pi)^d}~ \tilde{c}(p) e^{ip\cdot x} \cos\le( \sqrt{-p^2} ~ s\ri) \ ,
\ee
where $x$ denotes the coordinates on $\mathbb{R}^{1,d-1},$ $p^2 = -(p^0)^2 + (p^1)^2 + p^i p_i$ summing over $i \in \{2,...,d-1\},$ $s$ is an additional spatial coordinate such that $(x,s)$ covers $\mathbb{R}^{1,d}$ with the usual Minkowski metric, and
\be 
	\tilde{c}(p) = \int d^d x~ c(x) e^{-ip\cdot x}  \ ,
\ee
is the Fourier transform of $c(x).$ By construction $c_{ext}(x,s)$ satisfies 
 the massless wave equation on $\mathbb{R}^{1,d}$ ($\Box_x$ the wave operator on $\mathbb{R}^{1,d-1}$)
 \be 
	\le(\Box_x + {\pt^2 \ov \pt s^2}\ri) c_{ext}(x,s) = 0 \ ,
\ee
with a particular prescribed value, $c(x),$ on the timelike hypersurface $s=0$.

\begin{figure}[h]
\begin{centering}
	\includegraphics[width=2.5in]{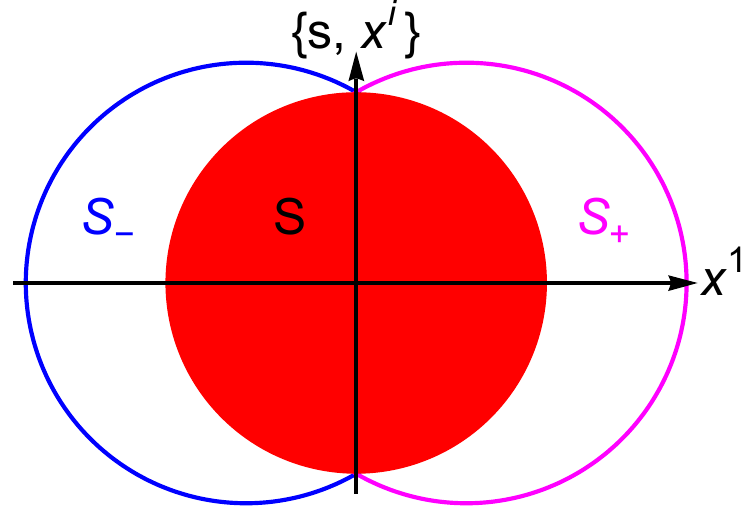}
\par\end{centering}
\caption{The $x^0=0$ slice of $\mathbb{R}^{1,d}$ appearing in the construction of $K(\hat R_- \cup \hat R_+)$. $K(\hat R_- \cup \hat R_+)$ is constructed by taking the union of all double cones based on $d$-spheres contained in $S_- \cup S_+$. Thus $K(\hat R_- \cup \hat R_+)$ must contain the double cone based on the red sphere $S$ in the figure. The restriction $K(\hat R_- \cup \hat R_+)$ to $s=0$ then contains new points that are not elements of $\hat R_- \cup \hat R_+$.}
\label{fig:constructK}
\end{figure}

We see that $c(x,0)$ vanishes for $x \in \hat{R}_- \cup \hat{R}_+$ (which we view as a subset of the $s=0$ hyperplane of $\mathbb{R}^{1,d}$). By the uniqueness theorem (see e.g.~theorem 7.2 of~\cite{WightmanAnFcn}) for solutions to the wave equation with prescribed behavior on a timelike curve or spatial sphere, we have that $c_{ext}(x,s) = 0$ in a certain larger region, denoted as $K\le(\hat{R}_- \cup \hat{R}_+\ri)$, on $\mathbb{R}^{1,d}$.  $K\le(\hat{R}_- \cup \hat{R}_+\ri)$ is defined to be the smallest subregion of $\mathbb{R}^{1,d}$ containing $\hat{R}_- \cup \hat{R}_+$ and all $(d+1)-$dimensional double cones spanned by timelike curves or based on (spatial) $d$-spheres that are contained in it.
Explicitly for the present case,
\be
	K\le(\hat{R}_- \cup \hat{R}_+\ri) = \bigcup\limits_{S \subset S_- \cup S_+} \hat{S} , 
\ee
with the union taken over all $d$-spheres $S$ contained in $S_- \cup S_+$ (see figure~\ref{fig:constructK}) where
\be 
	S_{\mp} = \le\{ (x^0,x^1,x^i,s) \in \mathbb{R}^{1,d} ~|~ x^0=0,~ (x^1 \pm b)^2 + x^i x_i + s^2 < a^2 \ri\}
\ee
and $\hat{S}$ is the domain of dependence in $\mathbb{R}^{1,d}$ of the sphere $S$. It turns out that in this simple case $K\le(\hat{R}_- \cup \hat{R}_+\ri) = \widehat{S_- \cup S_+} \subset \mathbb{R}^{1,d}$.
Projecting back to $s=0,$ one then concludes that $c_{ext}(x,0) = c(x)$ vanishes on 
\be
R^{\rm (vac)}_U \equiv K\le(\hat{R}_- \cup \hat{R}_+\ri) \biggr|_{s=0}  \ .
\ee
$R^{\rm (vac)}_U$ is larger than $\hat{R}_- \cup \hat{R}_+$ and, for $d>2,$ it turns out to be equal to the domain of dependence of $R_- \cup R_+$.
{For $d=2,$ $R^{\rm (vac)}_U$ is the region bounded by solid black and blue curves in figure~\ref{fig:fiT} (or equivalently, the blue region in figure~\ref{fig:overint} (a)).}

{Recalling that $\sY^{\rm (vac)}_{\hat{R}_- \cup \hat{R}_+}$ is a von Neumann algebra, it is equal to its double commutant. We have just shown that fields $\sO(x)$ for $x \in R^{\rm (vac)}_U$ commute with all operators in $\le(\sY^{\rm (vac)}_{\hat{R}_- \cup \hat{R}_+}\ri)'$ and thus are affiliated to $\sY^{\rm (vac)}_{\hat{R}_- \cup \hat{R}_+}.$ This establishes that $\sY^{\rm (vac)}_{\hat{R}_- \cup \hat{R}_+} \supsetneq \sY^{\rm (vac)}_{R^{\rm (vac)}_U},$ but, since $\hat{R}_- \cup \hat{R}_+ \subseteq R^{\rm (vac)}_U,$ we also have the reverse containment establishing that $\sY^{\rm (vac)}_{\hat{R}_- \cup \hat{R}_+} = \sY^{\rm (vac)}_{R^{\rm (vac)}_U}.$}

As an explicit example consider the case of $d=2.$ $R_U^{\rm (vac)}$ is then given by
\be
\begin{aligned}
	R_U^{\rm (vac)} = \le\{ (t,x) ~\bigg\lvert~
	 |t| < 
	\begin{cases}
		x + a + b, & x \in \le(-a-b, -b\ri) \\
		t_{vac}(x), & x \in \le(-b, b\ri) \\
		a + b - x, & x \in \le(b, a+b\ri) \\
	\end{cases}
	\ri\} \ ,
\end{aligned}
\ee
where
\be \label{eq:tVacAraki}
	t_{vac}(x) = \sqrt{x^2 + a^2 - b^2} \ ,
\ee
which clearly properly contains $\hat R_- \cup \hat R_+$.

\begin{figure}[h]
\begin{centering}
	\includegraphics[width=2.5in]{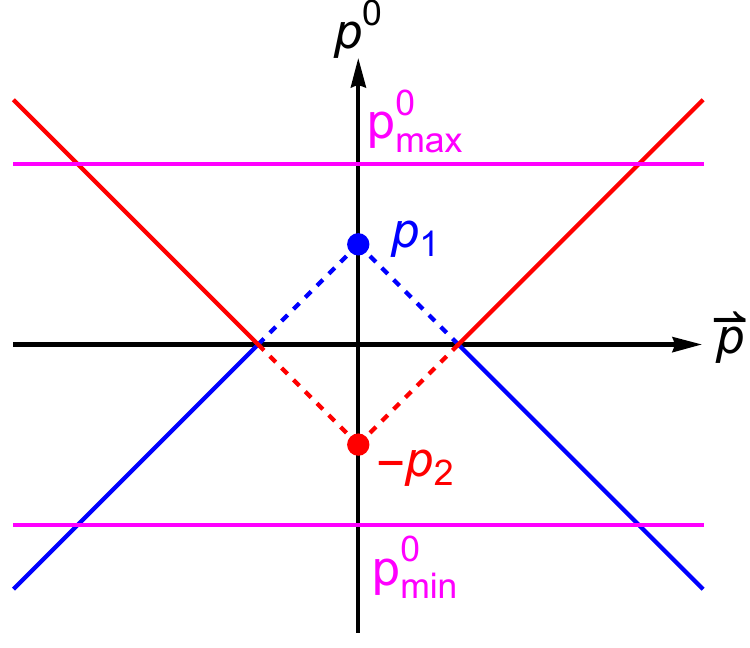}
\par\end{centering}
\caption{In the vacuum sector, $\tilde{c}(p)$, the Fourier transform of the commutator function~\eqref{eq:commFcn} can be suitably regulated to have compact support in $d$-momentum space yielding a smooth solution to the wave equation when extending to $(d+1)-$dimensional Minkowski space as in~\eqref{eq:extendFcn}. Specifically, since $\tilde{c}(p)$ is supported only in the region $\le(p_1 - \bar{V}_+\ri) \cup \le(-p_2 + \bar{V}_+\ri)$, putting a cutoff on energy $p^0_{\rm max}$ is sufficient to make the support compact. The support of $\tilde{c}(p)$ is contained between the solid red and blue lines in the figure so cutting off the energy to be between the magenta lines in the figure yields a compact subset of momentum space.}
\label{fig:pSpaceSupp}
\end{figure}

We have presented Araki's argument informally above in order to quickly reach the conclusion that $\sY^{\rm (vac)}_{\hat{R}_- \cup \hat{R}_+} = \sY^{\rm (vac)}_{R^{\rm (vac)}_U}$; however, we now comment on an important subtlety in the argument which shows that it is only valid in the vacuum sector. Specifically, the uniqueness theorems for the wave equation with prescribed data on a timelike surface only apply to smooth solutions, but~\eqref{eq:extendFcn} is not smooth on $\mathbb{R}^{1,d}$ 
as the cosine factor leads to exponential growth of the form $e^{\sqrt{p^2} |s|}$ at large spacelike momenta.
In the vacuum sector, the lack of smoothness of~\eqref{eq:extendFcn} can be circumvented by first
considering states $\ket{\Psi_{1,2}}$ that are only supported on a compact set of momenta in the forward lightcone ${V}_+$. 
More explicitly, we consider states $\ket{\Psi_{1,2}}$ that are only supported on the compact subregion $\le(p_{1,2} - {V}_+\ri) \cap {V}_+$ of momentum space for some $p_{1,2} \in {V}_+$.
As a result, $\tilde{c}(p)$ will be supported only on the subset of $\le(p_1 - {V}_+\ri) \cup \le(-p_2 + {V}_+\ri)$ since 
in the vacuum sector the momentum spectrum is contained in $V_+$.
We then further ``regularize''  $\tilde{c}(p)$ by multiplying it by some $\tilde{\chi}(p^0)$ of compact support in $p^0$,
yielding a function of compact support in $d-$momentum space.
See figure~\ref{fig:pSpaceSupp}.
Using this regulated function in~\eqref{eq:extendFcn} we obtain a smooth solution to the $(d+1)-$dimensional wave equation to which the uniqueness results can be applied. Araki then argues that the limit in which the regulating function $\tilde{\chi}(p^0)$ is removed and the restriction on the support of the involved states is lifted (i.e., taking $p_{1,2} \to \infty$) is well-controlled and one can therefore conclude that $c_{ext}(x,s)$ must vanish in $K\le(\hat{R}_- \cup \hat{R}_+\ri)$ as we obtained informally above. We direct the reader to~\cite{ArakiTT} for details.

An important consequence of this subtlety is that Araki's theorem does not apply when the spectrum of the theory does not lie in the forward light cone as, for example, happens when working about a thermal state. In the thermal sector of the theory, the Fourier transform $\tilde{c}(p)$ of the commutator function defined in~\eqref{eq:commFcn} will be supported on all of the $\mathbb{R}^{1,d-1}$ momentum space. In this case, one cannot apply Araki's regulation procedure to obtain a smooth extension of $c(x)$ to the higher-dimensional spacetime as the support of $\tilde{c}(p)$ at arbitrarily large spacelike momenta will always lead to a non-smooth extension due to the exponential growth of the cosine factor in~\eqref{eq:extendFcn} at such momenta.

\subsection{Overlapping regions in a thermal state} \label{app:overlapThermal}

In appendix~\ref{app:arakiVacArg} we summarized Araki's argument that shows that the algebra of two overlapping boundary double-cones contains operators localized at additional boundary points, outside of the union of original subregions. This argument assumed that the spectrum of the momentum operator lies in the forward lightcone, as is appropriate for the vacuum sector. This assumption fails for a thermal state and Araki's argument cannot be applied. However, using insights from holography, the algebra of two overlapping double-cones can still be seen to contain operators at additional boundary spacetime points.

As an illustrative example, let us consider the special case of two overlapping double cones $\hat R_{\pm}$ as in~\eqref{eq:defDoubleConesPM} (with $d=2$) on (1+1)-dimensional Minkowski space at finite temperature. We denote the associated thermal sector algebras by $\sY^{\rm (th)}_{\hat R_{\pm}}.$

Boundary operators about a thermal state on Minkowski space are dual to bulk fields propagating on the 
AdS-Rindler spacetime 
so $\sY^{\rm (th)}_{\hat R_{\pm}}$ is equal to the algebra of bulk fields in the corresponding causal domain $\fc_{R_{\pm}}$ on AdS-Rindler, i.e. $\sY^{\rm (th)}_{\hat R_{\pm}} = \widetilde \sM_{\fc_{R_{\pm}}}.$

Using additivity of the bulk theory we can compute the bulk spacetime region whose algebra of local bulk operators is equal to the additive algebra of the boundary diamonds. The intersection of this bulk region with the asymptotic boundary then gives a boundary subregion, which we denote by $R^{\rm (th)}_U$, whose corresponding operators are all in the additive algebra of the overlapping diamonds. Explicitly, we have $\sY^{\rm (th)}_{\hat{R}_{-}} \vee \sY^{\rm (th)}_{\hat{R}_{+}} = \widetilde{\sM}_{\fc_{R_-}} \vee \widetilde{\sM}_{\fc_{R_+}} = \widetilde{\sM}_{\widehat{\fc_{R_-} \cup \fc_{R_+}}} = \sY^{\rm (th)}_{R^{\rm (th)}_U}$. Thus $R^{\rm (th)}_U$ is the intersection of $\widehat{\fc_{R_-} \cup \fc_{R_+}}$ with the asymptotic boundary.

To compute $R^{\rm (th)}_U$ we need to find $\widehat{\fc_{R_-} \cup \fc_{R_+}}$ in the bulk. For a thermal boundary state on $(1+1)-$dimensional Minkowski space, the bulk dual has metric
\be 
	ds^2 = -f(r)dt^2 + f(r)^{-1} dr^2 + {r^2 \ov L_{AdS}^2} dx^2 = {L_{\rm AdS}^2 \ov w^2} \le(-\le(1-w^2\ri) \le({2\pi \ov \beta}\ri)^2 dt^2 + {dw^2 \ov 1-w^2} + \le({2\pi \ov \beta}\ri)^2 dx^2\ri) \ ,
\ee
with $f(r)=(r^2 - 4\pi^2L_{\rm AdS}^4\beta^{-2})L_{\rm AdS}^{-2}$, where $w \in (0,1)$ and $w=0$ is the asymptotic boundary and $w=1$ the AdS-Rindler horizon. The boundary of $\widehat{\fc_{R_-} \cup \fc_{R_+}}$ is found by firing orthogonal null geodesics from the boundary of $\fc_{R_-} \cup \fc_{R_+}$. In these coordinates, the null geodesics that reach the boundary satisfy (see e.g.~\cite{Sugishita})
\be 
	w(t) = {\sqrt{1-j^2} |\sinh\le({2\pi(t - t_0) \ov \beta}\ri)| \ov \sqrt{\cosh^2\le({2\pi(t - t_0) \ov \beta}\ri) - j^2 \sinh^2\le({2\pi(t - t_0) \ov \beta}\ri)}}, \qquad x(t) = x_0 + {\pi \ov \beta} \log\le({1+j \tanh\le({2\pi(t - t_0) \ov \beta}\ri) \ov 1 + j \tanh\le({2\pi(t - t_0) \ov \beta}\ri)}\ri) \ ,
\ee 
where $j \in (-1,1)$ is related to the spatial momentum and $(t_0,x_0)$ is the point at which the geodesic hits the asymptotic boundary. From these formulas, one can easily compute the intersection of the causal domain for $R_{\mp}$ with the (time-reflection symmetric) $t = 0$ slice. 

One finds 
\be 
	\fc_{R_{\mp}} = \le\{w < \cosh\le({2\pi(x \pm b) \ov \beta}\ri) \sqrt{\tanh^2 \le({2\pi a \ov \beta}\ri) - \tanh^2\le({2\pi(x \pm b) \ov \beta}\ri)},~ |x \pm b| < a \ri\} \ .
\ee
The boundaries of $\fc_{R_-}$ and $\fc_{R_+}$ intersect at the point
\be 
	p_* = (t_*,~ x_*,~ w_*) = \le(0,~0,~ {\sqrt{\cosh\le({4\pi a \ov \beta}\ri) - \cosh\le({4\pi b \ov \beta}\ri)} \ov \sqrt{2}\cosh\le({2\pi a \ov \beta}\ri)} \ri) \ .
\ee
The boundary of $\widehat{\fc_{R_-} \cup \fc_{R_+}}$ is comprised of a union of parts of the null surfaces obtained by taking ingoing orthogonal null rays fired from the part of $\pt \fc_{R_-}$ with $x<0$, ingoing orthogonal null rays fired from the part of $\pt \fc_{R_+}$ with $x>0$, and null rays fired from $p_*$.
The null rays from $p_*$ intersect the boundary in the surfaces defined by 
\be 
	t_{th}(x) = \pm {\beta \ov 2\pi} \text{arccosh} \le({\cosh \le({2\pi x \ov \beta}\ri) \ov \sqrt{1-w_*^2}}\ri) \ .
\ee
These surfaces define the upper/lower limits of the intersection of $\widehat{\fc_{R_-} \cup \fc_{R_+}}$ with the asymptotic boundary for $x \in (-b,b),$ so one obtains that the additive algebra $\sY^{\rm (th)}_{R_-} \vee \sY^{\rm (th)}_{R_+}$ 
is equal to the algebra, $\sY^{\rm (th)}_{R_U^{\rm (th)}},$ in the larger region
\be
\begin{aligned}
	R_U^{\rm (th)} = \le\{ (t,x) ~\bigg\lvert~
	 |t| < 
	\begin{cases}
		x + a + b, & x \in \le(-a-b, -b\ri) \\
		|t_{th}(x)|, & x \in \le(-b, b\ri) \\
		a + b - x, & x \in \le(b, a+b\ri) \\
	\end{cases}
	\ri\} \ .
\end{aligned}
\ee

Since $|t_{vac}(x)| < |t_{th}(x)|,$ one finds that there is a proper containment $R_U^{\rm (th)} \subsetneq R_U^{\rm (vac)}$ of extended boundary spacetime subregions. This shows that the result of combining algebras of two overlapping diamond-shaped boundary subregions results in a strictly larger geometric extension of the region in the vacuum case compared to the thermal case.
An example of the difference in extensions is shown in figure~\ref{fig:fiT}.

An interesting phenomenon is that the discrepancy between the thermal and vacuum extensions, $R_U^{\rm (th)}$ and $R_U^{\rm (vac)},$ becomes more pronounced as the temperature is increased. Fixing the width of the diamonds, $2a,$ and the separation of their centres, $2b,$ we can consider the low and high temperature limits. At low temperature ($\beta \to \infty$) one can easily check that $t_{th}(x) = t_{vac}(x) + O(\beta^{-2}),$ demonstrating that the thermal and vacuum extensions agree in this limit. This is to be expected since the algebras considered only probe scales much smaller than the thermal scale and thus cannot distinguish between the vacuum and thermal states. Conversely, one can show that, for $x \in (-b,b),$ $t_{th}(x) = a-b+|x| + O(\beta e^{-1/\beta})$ in the high temperature $(\beta \to 0)$ limit. Thus at high temperature the extension, $R_U^{\rm (th)},$ reduces to the original subregion $\hat{R}_- \cup \hat{R}_+$ and the additive algebra does not probe any additional boundary points. 
At high temperature, there is also a large discrepancy between the vacuum extension $R_U^{\rm (vac)}$ and thermal extension $R_U^{\rm (th)}.$ This can be easily understood since the fixed-size boundary subregions now probe scales much larger than the thermal scale so the corresponding algebras are sensitive to distinctions between the thermal and vacuum states.

We emphasize that our result could not have been obtained if Araki's argument applied. His argument would have established that the additive algebra of $\hat{R}_-$ and $\hat{R}_+$ is equivalent to the algebra of $R_U^{\rm (vac)}$ while we instead find the additive algebra is equivalent to algebra of strictly smaller subregion $R_U^{\rm (th)}.$ This example demonstrates that the assumption of spectrum in the forward lightcone is crucial to Araki's argument. In particular, for the thermal case in $(1+1)-$dimensions we find that $\tilde{c}(p)$ decays exponentially (but never vanishes) as $e^{-{\beta \ov 4} \sqrt{p^2}}$ at large spacelike momenta~\cite{Morrison:2014jha, longPaper} and hence the commutator function can only be extended as in~\eqref{eq:extendFcn} smoothly up to $s = {\beta \ov 4}.$ We may think of this limit on how far we can extend boundary commutator functions into the bulk as a signature that the bulk dual has a horizon.

\section{Relation between $S_E$ and  $S$} \label{app:SEcontS}

In this appendix we discuss the relation between the $S_E$ and $S$. We first prove that 
$S_E \supseteq S$, and then show that for empty AdS$_3$, $S_E = S$.

\subsection{Proof of $S_E \supseteq S$}

The main argument of the proof is the same as one given in~\cite{May:2019odp} for $S \subseteq \hat \fb_{V_1 \cup V_2}.$ The only new ingredient is to use a time-reversal of that argument to also conclude that $S \subseteq \hat \fb_{W_1 \cup W_2}$. Explicitly we want to prove
\be 
	S_E \equiv \hat \fb_{V_1 \cup V_2} \cap \hat \fb_{W_1 \cup W_2} \supseteq S \equiv J^+(\hat \fb_{V_1})\cap J^+(\hat \fb_{V_2})\cap J^-(\hat \fb_{W_1})\cap J^-(\hat \fb_{W_2}) \ .
\ee
In order for this to be true we need to assume that the bulk satisfies the null curvature condition and that the boundary state is pure.

{\bf Proof:} We assume $S$ is non-empty since if $S$ is empty, the containment is trivially true. This implies that the entanglement wedge of $V \equiv V_1 \cup V_2$ is connected (by the usual CWT of~\cite{May:2019yxi}). Since the boundary state is pure, the RT surface is the same for both $X \equiv X_1 \cup X_2 = (V_1 \cup V_2)'$ and $V$ and thus the entanglement wedges are bulk causal complements: $\hat \fb_{V_1 \cup V_2} = (\hat \fb_{X_1 \cup X_2})'.$ Because the entanglement wedge of $V$ is connected, the entanglement wedge of $X$ must then be disconnected and we have $\hat \fb_X = \hat \fb_{X_1} \vee \hat \fb_{X_2}.$

The entanglement wedge of $X$ divides the bulk into four disjoint regions, $\hat \fb_{X}$ itself, its causal future, causal past, and causal complement. Consider a bulk point $p \in S.$ Note that a bulk operator at $p$ cannot commute with all operators in $\hat \fb_{V_1}$ or in $\hat \fb_{V_2}.$
Thus $p$ cannot be in $\hat \fb_{X}$ as then $\sX_{X} = \widetilde{\sM}_{\fb_{X}}$ would contain an operator that does not commute with all operators in $\sX_{V_1}$ and $\sX_{V_2}$ violating boundary causality. 

Similarly, $p$ cannot be to the causal past of $\hat \fb_{X}$. For purposes of contradiction, suppose that $p \in J^-(\hat \fb_{X}),$ then there exists a bulk past-directed causal curve $c_X$ from $\hat \fb_X$ to $p.$ But since $p \in J^+(\hat \fb_{V_1})$ there also exists a bulk future-directed causal curve $c_{V_1}$ from $\hat \fb_{V_1}$ to $p.$ Thus, the bulk curve $c_{V_1} \cup \overline{c_{X}}$ is a future-directed bulk causal curve from $\hat \fb_{V_1}$ to $\hat \fb_{X}.$\footnote{Here the overbar indicates the time-reversal of the bulk curve.} However, no such bulk curve can exist since $\hat \fb_{V_1} \subsetneq \hat \fb_{V} = (\hat \fb_{X})'$ is spacelike to $\hat \fb_{X}.$ 
Thus, any $p \in S$ cannot lie in $J^-(\hat \fb_{X}).$ 

It is also not possible for $p$ to lie to the future of $\hat \fb_{X}.$ To see this, suppose, for the purpose of contradiction, that $p \in J^+(\hat \fb_{X} = \hat \fb_{X_1} \vee \hat \fb_{X_2}),$ then since $p \in S \subsetneq J^-(\hat \fb_{W_1}) \cap J^-(\hat \fb_{W_2})$ there must exist bulk causal curves either from $\hat \fb_{X_1}$ or $\hat \fb_{X_2}$ to both $\hat \fb_{W_2}$ and $\hat \fb_{W_1}$. Without loss of generality, assume the labeling of regions is such that there are bulk causal curves from $\hat \fb_{X_1}$ to both $\hat \fb_{W_1}$ and $\hat \fb_{W_2}$. This implies that not all operators in $\sX_{X_1} = \widetilde{\sM}_{\fb_{X_1}}$ will commute with all operators in $\sX_{W_1} = \widetilde{\sM}_{\fb_{W_1}}$ nor with all operators in $\sX_{W_2} = \widetilde{\sM}_{\fb_{W_2}}$. However, this contradicts boundary causality since $X_1$ is spacelike to one of $W_1$ or $W_2$. So $p$ cannot lie in the causal future of $\hat \fb_{X}.$

We therefore conclude that $p \in (\hat \fb_{X_1 \cup X_2})'$, so we have shown that $p \in S$ implies $p \in \hat \fb_{V_1 \cup V_2}.$ 

The time reversal of the above argument with $V_{1,2} \to W_{1,2}$ and $X_{1,2} \to Y_{1,2}$ establishes that $p \in S$ implies $p \in \hat \fb_{W_1 \cup W_2}.$ We define $W\equiv W_1 \cup W_2$ and $Y \equiv W' = Y_1 \cup Y_2.$ Since $S\neq\emptyset$ the usual CWT~\cite{May:2019yxi} implies $\fb_W$ is connected. Since the boundary state is pure, its causal complement is $\hat \fb_Y$ and we have $\hat \fb_Y = \hat \fb_{Y_1} \vee \hat \fb_{Y_2}$. 

Consider $p\in S.$ We cannot have $p \in \hat \fb_Y$ since $\sX_Y = \widetilde{\sM}_{\fb_Y}$ and the existence of such a bulk operator at $p$ would imply the existence of a boundary operator in $Y$ that does not commute with all operators in $W$ (since $p \in J^-(\hat \fb_{W_1})\cap J^-(\hat \fb_{W_2})$) violating boundary causality. Similarly we cannot have $p \in J^+(\hat \fb_Y)$ as this would imply the existence of a past-directed bulk causal curve from $\hat \fb_{W_1}$ to $\hat \fb_Y$ passing through $p,$ violating $\hat \fb_{W_1} \subsetneq (\hat \fb_{Y})'.$ We also cannot have $p \in J^-(\hat \fb_{Y})$ since then there would exist a past-directed bulk causal curve from either $\hat \fb_{Y_1}$ or $\hat \fb_{Y_2}$ to both $\hat \fb_{V_1}$ and $\hat \fb_{V_2}$ implying for example, by the duality $\sX_{Y_{1,2}} = \widetilde{\sM}_{\fb_{Y_{1,2}}},$ that there are operators in $Y_{1}$ that do not commute with all operators in $V_1$ nor with all operators in $V_2.$ This contradicts boundary causality since $Y_1$ and $V_2$ are spacelike separated.

We therefore conclude that $p \in (\hat \fb_{Y_1 \cup Y_2})'$, so we have shown that $p \in S$ implies $p \in \hat \fb_{W_1 \cup W_2}.$ 

Putting these arguments together we have the implication $p \in S \Rightarrow p \in (\hat \fb_{X})'\cap (\hat \fb_{Y})'.$ For a pure state we have the equivalences $(\hat \fb_{X})' = \hat \fb_V$ and $(\hat \fb_{Y})' = \hat \fb_W$ so we have shown $p \in S \Rightarrow p \in \hat \fb_V\cap \hat \fb_W \equiv S_E.$ Thus, $S_E \supseteq S$ as claimed.

\subsection{$S_E =S$ in empty AdS$_3$}

\begin{figure}[!h]
\begin{center}
\includegraphics[width=6.5cm]{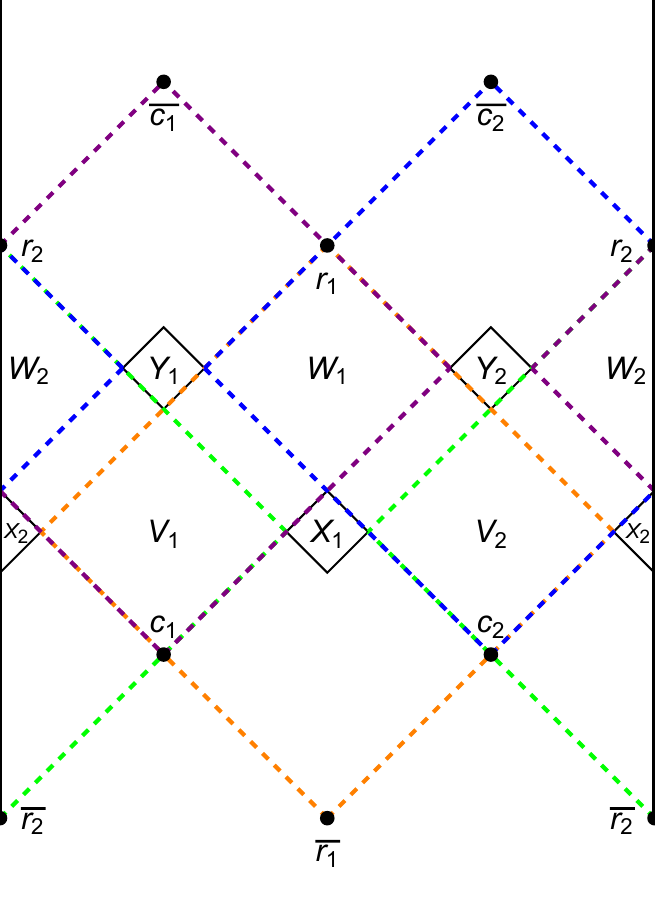}
\caption{ The points $\bar r_{1,2},~ \bar c_{1,2}$ used to construct $(X_{1,2})'$ and $(Y_{1,2})'.$ $(X_1)'$ is shown with green dashed lines, $(X_2)'$ with orange dashed lines, $(Y_1)'$ with blue dashed lines and $(Y_2)'$ with purple dashed lines.
}
\label{fig:c1c2r1r2bar}
\end{center}
\end{figure}

We now argue that, in pure AdS$_3$, $S_E = S$.  Consider again a configuration, such as that depicted in figure~\ref{fig:connWedge}, for which $V_{1,2} \neq \emptyset,$ but no boundary scattering region exists. Recall that we defined single diamonds $X_{1,2}$ such that $X_1 \cup X_2 = (V_1 \cup V_2)'.$ Their causal complements are characterized by 
\be 
	\le(X_1\ri)' = \hat{J}^-(r_2) \cap \hat{J}^+(\bar r_2), \qquad \le(X_2\ri)' = \hat{J}^-(r_1) \cap \hat{J}^+(\bar r_1) \ ,
\ee
where the points $\bar r_{1,2}$ are shown in figure~\ref{fig:c1c2r1r2bar}. Similarly we have $Y_1 \cup Y_2 = (W_1 \cup W_2)',$ with each of $Y_{1,2}$ being a single diamond. Their causal complements are 
\be 
	\le(Y_1\ri)' = \hat{J}^+(c_2) \cap \hat{J}^-(\bar c_2), \qquad \le(Y_2\ri)' = \hat{J}^+(c_1) \cap \hat{J}^-(\bar c_1) \ .
\ee
The points $\bar c_{1,2}$ are shown in figure~\ref{fig:c1c2r1r2bar}. Note that the causal domains (which are equal to entanglement wedges in this case) of these regions can be similarly described by replacing the boundary causal past/future $\hat{J}^{\mp}$ with the bulk causal past/future $J^{\mp}$, i.e. we have
\be 
	\hat \fb_{(X_1)'} = \hat \fc_{(X_1)'} = J^-(r_2) \cap J^+(\bar r_2), \qquad \hat \fb_{(X_2)'} = \hat \fc_{(X_2)'} = J^-(r_1) \cap J^+(\bar r_1) 
\ee
\be 
	\hat \fb_{(Y_1)'} = \hat \fc_{(Y_1)'} = J^+(c_2) \cap J^-(\bar c_2), \qquad \hat \fb_{(Y_2)'} = \hat \fc_{(Y_2)'} = J^+(c_1) \cap J^-(\bar c_1) \ . 
\ee

Consider the bulk subregions
\be 
	\fa_V \equiv \le(\hat \fc_{X_1} \cup \hat \fc_{X_2}\ri)' = \le(\hat \fc_{X_1}\ri)' \cap \le(\hat \fc_{X_2}\ri)' = \hat \fc_{(X_1)'} \cap \hat \fc_{(X_2)'} \ ,
\ee
\be 
	\fa_W \equiv \le(\hat \fc_{Y_1} \cup \hat \fc_{Y_2}\ri)' = \le(\hat \fc_{Y_1}\ri)' \cap \le(\hat \fc_{Y_2}\ri)' = \hat \fc_{(Y_1)'} \cap \hat \fc_{(Y_2)'} \ .
\ee
The final equalities can be seen to follow from the fact that the causal and entanglement wedges coincide in this case and the identification of the entanglement wedges respects taking causal complements on the boundary.

Since $\bar{r_1},~\bar{r_2}$ are each to the past of $c_1$ and $c_2,$ while $\bar{c_1},~\bar{c_2}$ are each to the future of $r_1$ and $r_2,$ one can easily see that the intersection of $\fa_V$ with $\fa_W$ is the region
\be \label{eq:intAs}
	\fa_V \cap \fa_W = J^+(c_1) \cap J^+(c_2) \cap J^-(r_1) \cap J^-(r_2) = S_0 \ .
\ee

The entanglement and causal wedges of boundary diamonds coincide in pure AdS so we have that
\be 
	J^+(\hat \fb_{V_i}) = J^+(c_i), \qquad J^-(\hat \fb_{W_i}) = J^-(r_i) \ ,
\ee
for $i = 1,2.$ Thus the RHS of~\eqref{eq:intAs} is clearly equal to the generalized bulk scattering region, $S,$ and we have
\be \label{eq:scattEqEWVac}
	S \equiv J^+(\hat \fb_{V_1})\cap J^+(\hat \fb_{V_2})\cap J^-(\hat \fb_{W_1})\cap J^-(\hat \fb_{W_2}) = \fa_V \cap \fa_W \ .
\ee
We therefore see that the generalized scattering region in pure AdS$_3$ can always be written as an intersection of four causal wedges. When $V_1 \cup V_2$ and $W_1 \cup W_2$ are in the connected phase, the subregions $\fa_V$ and $\fa_W,$ respectively, correspond to their entanglement wedges, so in this case, in addition to being the intersection of four causal wedges, $S$ is also the intersection of two entanglement wedges.

To see this, consider the case of a connected entanglement wedge for $V_1 \cup V_2$. The entanglement wedge of $V_1 \cup V_2$ and that of its complement $X_1 \cup X_2$ are bulk causal complements. 
Since we are assuming that the entanglement wedge of $V_1 \cup V_2$ is connected we then see that the entanglement wedge of $X_1 \cup X_2$ must be disconnected. One can in fact show that the entanglement wedge of $X_1 \cup X_2$ is equal to its causal wedge. As $V_1 \cup V_2$ and $W_1 \cup W_2$ are always in the same phase, we can also conclude that the entanglement wedge of $Y_1 \cup Y_2$ is equal to its causal wedge.

Thus, in this connected phase we have that
\be 
	\fa_V \equiv (\hat \fc_{X_1} \cup \hat \fc_{X_2})' = (\hat \fb_{X_1 \cup X_2})' = \hat \fb_{V_1 \cup V_2} \ ,
\ee
\be 
	\fa_W \equiv (\hat \fc_{Y_1} \cup \hat \fc_{Y_2})' = (\hat \fb_{Y_1 \cup Y_2})' = \hat \fb_{W_1 \cup W_2} \ .
\ee

So from~\eqref{eq:scattEqEWVac}, when $V_1 \cup V_2$ is in the connected phase we have that 
\be \label{eq:scattRegIsEWIntNonEmpty}
	S = \hat \fb_{V_1 \cup V_2} \cap \hat \fb_{W_1 \cup W_2} \le( \neq \emptyset \ri) \ .
\ee
Alternatively, when $V_1 \cup V_2$ is in the disconnected phase both sides of the above relation are empty, so the relation trivially holds.\footnote{To see this, recall that we have already shown that when $V_1 \cup V_2$ is in the disconnected phase $S$ is empty. In this phase the entanglement and causal wedges for $V_1 \cup V_2$ (and for $W_1 \cup W_2$) agree so the RHS of~\eqref{eq:scattRegIsEWInt} is $\le(\hat \fc_{V_1} \cup \hat \fc_{V_2}\ri) \cap \le(\hat \fc_{W_1} \cup \hat \fc_{W_2}\ri)$. This can only be non-empty if at least one of $\hat \fc_{V_i} \cap \hat \fc_{W_j},$ for $i,j \in \{1,2\},$ is non-empty. However this would imply that there is a bulk point, $p,$ in the future of the past-most point in $W_j$ and past of the future-most point in $V_i.$ A curve constructed from the union of a bulk causal curve from the past-most point of $W_j$ to $p$ and a bulk causal curve from $p$ to the future-most point of $V_i$ would then be a bulk causal curve between two spacelike separated boundary points. However, by the Gao-Wald theorem~\cite{Gao:2000ga}, no such bulk curve exists. } Thus, about the vacuum state on $\mathbb{R} \times S^1$ the relation
\be \label{eq:scattRegIsEWInt}
	S \equiv J^+(\hat \fb_{V_1})\cap J^+(\hat \fb_{V_2})\cap J^-(\hat \fb_{W_1})\cap J^-(\hat \fb_{W_2}) = \hat \fb_{V_1 \cup V_2} \cap \hat \fb_{W_1 \cup W_2} \equiv S_E\ .
\ee
always holds.

\section{The GCWT in empty AdS$_3$} \label{app:vacGCWT}

In this appendix, we prove the GCWT in a special case: empty AdS$_3$. 
In this case, $S_E = S$ (see Appendix~\ref{app:SEcontS} for a proof), and hence in the forward direction 
the GCWT reduces to the CWT.   
While in this case the GCWT amounts to the CWT and its converse, the reformulation using $S_E$ highlights the equivalence between the
existence of a generalized bulk scattering region and of operators that are {\it not obtained additively} from $V_1$ and $V_2$.

Recall that the CWT is the statement
\be 
	S \text{ has non-empty interior } \Rightarrow \fb_{V_1 \cup V_2} \text{ is connected.}
\ee
The converse is then
\be
	\fb_{V_1 \cup V_2} \text{ is connected } \Rightarrow S \text{ has non-empty interior. }
\ee

Recall from our earlier discussion (see Fig.~\ref{fig:connWedge}) that $V_1$ ($V_2$) is a domain of dependence of an interval $\sigma_1$ ($\sigma_2$).  We denote the `left' endpoint of $\sigma_i$ by $x_i$ and the `right' endpoint by $y_i$ for $i\in \{1,2\}$.

Non-existence of a boundary scattering region, but non-emptiness of $V_1$ and $V_2,$  implies that $V_1 \cup V_2$ can be completely described by the four (pairwise) spacelike separated points $x_1,y_1,x_2,y_2.$ For any set of all mutually spacelike separated points on the Lorentzian cylinder there exists a Minkowski coordinate patch that contains all of those points. As is well known, on (1+1)-dimensional Minkowski space we may use a conformal transformation to bring any four spacelike separated points to a constant time-slice and place those points at spatial coordinates $0,\eta,1,\infty,$ respectively, where $\eta \in (0,1)$ is the cross ratio --  the only conformal invariant that can be built by four points (in two dimensions). It is then clear that, up to a conformal transformation, any configuration of input and output points with no boundary scattering region and non-empty $V_{1,2}$ (these conditions are defined by causal relations and thus are conformally invariant) can be characterized by a single number, the cross ratio ($\eta$) of the four endpoints of $V_1$ and $V_2.$ Importantly, about the vacuum state, $\eta$ also determines the phase of the entanglement wedge for $V_1 \cup V_2:$ if $\eta < \ha$ the entanglement wedge is disconnected, while if $\eta > \ha$ the entanglement wedge is connected. One can show that the cross ratio characterizing $V_1 \cup V_2$ is equal to that for $W_1 \cup W_2,$ thus, about the vacuum state, $V_1 \cup V_2$ and $W_1 \cup W_2$ are always in the same phase.

To proceed, we introduce coordinates covering global AdS$_3,$ $(t,\rho,\theta),$ with metric
\be 
	ds^2 = {l^2 \ov \cos^2\rho} \le(-dt^2 + d\rho^2 + \sin^2\rho ~d\theta^2 \ri) \ ,
\ee
where $\rho = 0$ is the centre of rotational symmetry and $\rho = {\pi \ov 2}$ is the asymptotic boundary. This boundary is a Lorentzian cylinder that we describe with global coordinates $(t,\theta).$ Without loss of generality we can perform a global time translation and rotation to place all of the endpoints of the boundary intervals $V_{1,2}$, i.e. $x_1,y_1,x_2,y_2,$ in the range $|t \pm \theta| < \pi.$ This subregion of the Lorentzian cylinder can be described in Minkowski coordinates. We use null Minkowski coordinates $x^{\pm} \in \mathbb{R}^2$ where the Minkowski metric is $ds^2_M = -dx^+dx^-.$ The Minkowski coordinates in this patch are related to the global coordinates on the cylinder by 
\be \label{eq:minkEmbed}
	x^{\pm} = \tan \le( {t \pm \theta \ov 2}\ri),
\ee
so we see that the global and Minkowski metrics are related by
\be 
	ds^2_G = -d(t-\theta)d(t + \theta) = - \le(2 \ov 1 + (x^+)^2\ri)\le(2 \ov 1 + (x^-)^2\ri) dx^+dx^- = \le(2 \ov 1 + (x^+)^2\ri)\le(2 \ov 1 + (x^-)^2\ri) ds^2_M \ .
\ee

We perform a conformal transformation that preserves this Minkowski patch on the cylinder and brings the four points describing $V_1 \cup V_2$ to the following locations
\be \label{eq:vEndpoints}
	x_1: \le(t=0,~ \theta = \pi - { w \ov 2}\ri),~ y_1: \le(t=0,~ \theta = -\pi + {w \ov 2}\ri),~  x_2: \le(t=0,~ \theta = -{w \ov 2}\ri),~ y_2: \le(t=0,~ \theta = {w \ov 2}\ri) ,
\ee
where $w \in (0,\pi)$ is the total angular width of each of $V_1$ and $V_2$. Using~\eqref{eq:minkEmbed} the cross ratio of this configuration is computed to be
\be \label{eq:crossRatW}
	\eta = {\le(\tan\le({\pi - w/2 \ov 2}\ri) - \tan\le({-\pi + w/2 \ov 2} \ri)\ri)\le(\tan\le({- w/2 \ov 2}\ri) - \tan\le({ w/2 \ov 2} \ri)\ri) \ov \le(\tan\le({\pi - w/2 \ov 2}\ri) - \tan\le({-w/2 \ov 4} \ri)\ri)\le(\tan\le({-\pi + w/2 \ov 2}\ri) - \tan\le({w/2 \ov 2} \ri)\ri)} = \sin^2\le({w \ov 2}\ri) \ .
\ee
Notice that $\eta$ ranges from zero to one as we increase the parameter $w$ from zero to $\pi.$ We therefore see that, up to a conformal transformation, the specific choices~\eqref{eq:vEndpoints} parametrize all possible two-to-two scattering configurations with no boundary scattering region and non-empty $V_{1,2}.$ 

With the choice of endpoints of $V_{1,2}$ given in~\eqref{eq:vEndpoints} the corresponding configuration of input and output points is
\bega \label{eq:specialInpOutPts}
	c_1: \le(t = -{w\ov 2}, \theta = \pi\ri),~ c_2: \le(t = -{w\ov 2}, \theta = 0\ri), \\
	 r_1: \le(t = {\pi + w\ov 2}, \theta = -{\pi \ov 2}\ri),~ r_2: \le(t = {\pi + w\ov 2}, \theta = {\pi \ov 2}\ri) \ .
\end{gather} 

Recall that, since the entanglement and causal wedges coincide for a boundary interval in pure AdS$_3$ we have $J^+(\hat \fb_{V_i}) = J^+(c_i)$ and $J^-(\hat \fb_{W_i}) = J^-(r_i).$ Denote the constant global time slice of pure AdS$_3$ at time $t$ by $\Sigma_t.$ By explicit computation, one can show that $(J^+(c_1)\cap J^+(c_2))\cap \Sigma_t$ is empty whenever $t < {\pi - w \ov 2}.$ Similarly, one can show that $(J^-(r_1)\cap J^-(r_2))\cap \Sigma_t$ is empty for $t > {w \ov 2}.$ Thus, when ${w \ov 2} < {\pi - w \ov 2} \Leftrightarrow w < {\pi \ov 2}$ there cannot possibly be a bulk scattering region as there is no time at which $J^+(c_1)\cap J^+(c_2)$ and $J^-(r_1)\cap J^-(r_2)$ are both non-empty. 

This proves the CWT in vacuum. For $(0<)~w < {\pi \ov 2},$ by~\eqref{eq:crossRatW} we have $\eta < \ha$ and thus the entanglement wedge of $V_1 \cup V_2$ is disconnected. But we have just argued that $w < {\pi \ov 2}$ implies that the scattering region must be empty. So a disconnected entanglement wedge for $V_1 \cup V_2$ implies an empty scattering region. Taking the negation we have the CWT: a non-empty scattering region implies a connected entanglement wedge.

To show the converse of the CWT in vacuum now consider a configuration with $w > {\pi \ov 2}$ so that the entanglement wedge of $V_1 \cup V_2$ is connected. Consider the fixed point, $c_t = (t,0,0),$ of the bulk rotation symmetry at time $t.$ 
For ${\pi -w \ov 2} < t < {w \ov 2}$ one can easily see that $c_t \in S_0,$ i.e. all such $c_t$ are in the scattering region. In fact, any bulk point in the timelike envelope of the curve $\gamma(t) = (t,0,0),~ t \in \le({\pi -w \ov 2}, {w \ov 2}\ri)$ is in the scattering region, so the scattering region has non-empty interior. Thus we have shown that the converse of the CWT holds in vacuum: a connected entanglement wedge implies that there is a bulk scattering region with non-empty interior.

\end{document}